\documentclass[iop]{emulateapj}

\usepackage{times,graphicx}
\usepackage[backref,breaklinks,colorlinks,citecolor=blue]{hyperref}
\usepackage[all]{hypcap}
\usepackage{amssymb,amsmath}
\usepackage{aas_macros}
\usepackage{natbib}
\usepackage{color}
\usepackage{tabularx}
\usepackage{bm}
\usepackage{array}

\newcommand{\kepler}{\emph{Kepler}}

\newcommand{\thetae}{\theta_{\mathrm{E}}}
\newcommand{\re}{r_{\mathrm{E}}}

\newcommand{\pirel}{\pi_{\rm rel}}
\newcommand{\dl}{D_{\mathrm{l}}}
\newcommand{\ds}{D_{\mathrm{s}}}

\newcommand{\powhalf}[1]{{#1}^{\frac{1}{2}}}

\newcommand{\mpowhalf}[1]{{#1}^{-\frac{1}{2}}}

\newcommand{\half}{\frac{1}{2}}
\newcommand{\msun}{M_{\odot}}

\newcommand{\dd}{\mathrm{d}}

\newcommand{\phalf}{^{\frac{1}{2}}}

\newcommand{\mfracpow}[3]{#1^{-\frac{#2}{#3}}}

\newcommand{\rabbdot}{\dot{\bm{r}}_{\rm ABb}}
\newcommand{\rabb}{\bm{r}_{\rm ABb}}
\newcommand{\rabdot}{\dot{\bm{r}}_{\rm Ab}}
\newcommand{\rab}{\bm{r}_{\rm Ab}}

\newcolumntype{C}[1]{>{\centering\let\newline\\\arraybackslash\hspace{0pt}}m{#1}}

\begin{document}

\title{Caustic Structures and Detectability of Circumbinary Planets in Microlensing}
\shorttitle{Microlensing \& Circumbinary Planets}
\shortauthors{Luhn, Penny \& Gaudi}

\author{Jacob K. Luhn\altaffilmark{1,2}, Matthew T. Penny\altaffilmark{1,3}, B. Scott Gaudi\altaffilmark{1}}
\altaffiltext{1}{Department of Astronomy, Ohio State University, 140 West 18th Avenue, Columbus, OH 43210, USA}

\altaffiltext{2}{Department of Astronomy \& Astrophysics, The Pennsylvania State University, 525 Davey Lab, University Park, PA 16802, USA}

\altaffiltext{3}{NASA Sagan Fellow}
\email{jluhn@psu.edu}

\begin{abstract}
Recent discoveries of circumbinary planets in \kepler\ data show that there is a viable channel of planet formation around binary main sequence stars. Motivated by these discoveries, we have investigated the caustic structures and detectability of circumbinary planets in microlensing events. We have produced a suite of animations of caustics as a function of the projected separation and angle of the binary host to efficiently explore caustic structures over the entire circumbinary parameter space. Aided by these animations, we have derived a semi-empirical analytic expression for the location of planetary caustics, which are displaced in circumbinary lenses relative to those of planets with a single host. We have used this expression to show that the dominant source of caustic motion will be due to the planet's orbital motion and not that of the binary star. Finally, we estimate the fraction of circumbinary microlensing events that are recognizable as such to be significant ($5$--$50$~percent) for binary projected separations in the range $0.1$--$0.5$ in units of Einstein radii.

\end{abstract}


\section{Introduction}\label{intro}

Using data from the \kepler\ spacecraft, nine circumbinary planet systems have been discovered to date~\citep[e.g.,][]{Doyle2011,Welsh2015}, demonstrating that there exists a viable channel of planet formation around main sequence binary stars. Indeed,~\citet{Armstrong2014} have estimated the abundance of circumbinary planets to be 10~percent, which is comparable to the abundance of planets around single stars. However, because the transit technique used by \kepler\ is sensitive to close-orbiting planets around close-orbiting binary stars, the circumbinaries discovered by \kepler\ have separations of only 1 AU or less. Finding planets in wider orbits or circumbinary planets around wider binaries requires a different technique. There is one known example of a planet on a ${\sim}100$~yr orbit around a millisecond pulsar-white dwarf binary~\citep{Thorsett1999}, but this object is thought to have been captured during a close encounter with another globular cluster star~\citep{Sigurdsson1993}. Evidence for planetary companions to post common envelope binaries have been claimed based on eclipse timing data, but the cause of the timing variations has yet to be conclusively proved~\citep[see, e.g., the discussion in][]{Parsons2014} and it is possible that another mechanism is responsible for the timing variations~\citep[e.g.,][]{Applegate1992}. A search for circumbinary planets on wider orbits around main sequence stars could shed light on whether they form in-situ~\citep{Meschiari2014} or migrated inwards from a formation site further out~\citep{Kley2014}.

Gravitational microlensing is sensitive to planets with projected separations within a factor of ${\sim} 2$ of the Einstein ring radius \citep[see, e.g.,][]{Gaudi2012}, 
\begin{equation}
\label{einsteinring}
\re = \dl \thetae
\end{equation}
 where $\dl$ is the distance to the lens and
\begin{equation}
\thetae = \sqrt{\kappa M \pirel}, 
\label{thetae}
\end{equation}
is the angular Einstein radius, where $M$ is the mass of the host lens, $\pirel=\text{AU}(\dl^{-1}-\ds^{-1})$ is the relative lens-source parallax, with $\ds$ being the source distance; finally $\kappa=4G/c^2=8.144$~mas~$\msun^{-1}$ is a constant with $G$ and $c$ having their usual meaning. The typical scale of the Einstein ring is $2$--$3$~AU. Microlensing is also sensitive to the binarity of lenses with a larger range of projected separations, ${\sim}0.1$--$10$~$\re$. Because microlensing does not rely on the detection of light from the host or planet, microlensing surveys typically probe a wider range of host masses than other techniques. This means that microlensing is potentially sensitive to both the planets and the host binaries of a wide range of circumbinary systems that might be inaccessible to other planet finding techniques. However, there has been little work to date on circumbinary planet microlensing, and it is not yet clear the extent to which a binary lens star might suppress or enhance the detectability of planets in a circumbinary system, or how detectable the binary nature of the host star is. 

\begin{figure}
\includegraphics[width=\columnwidth]{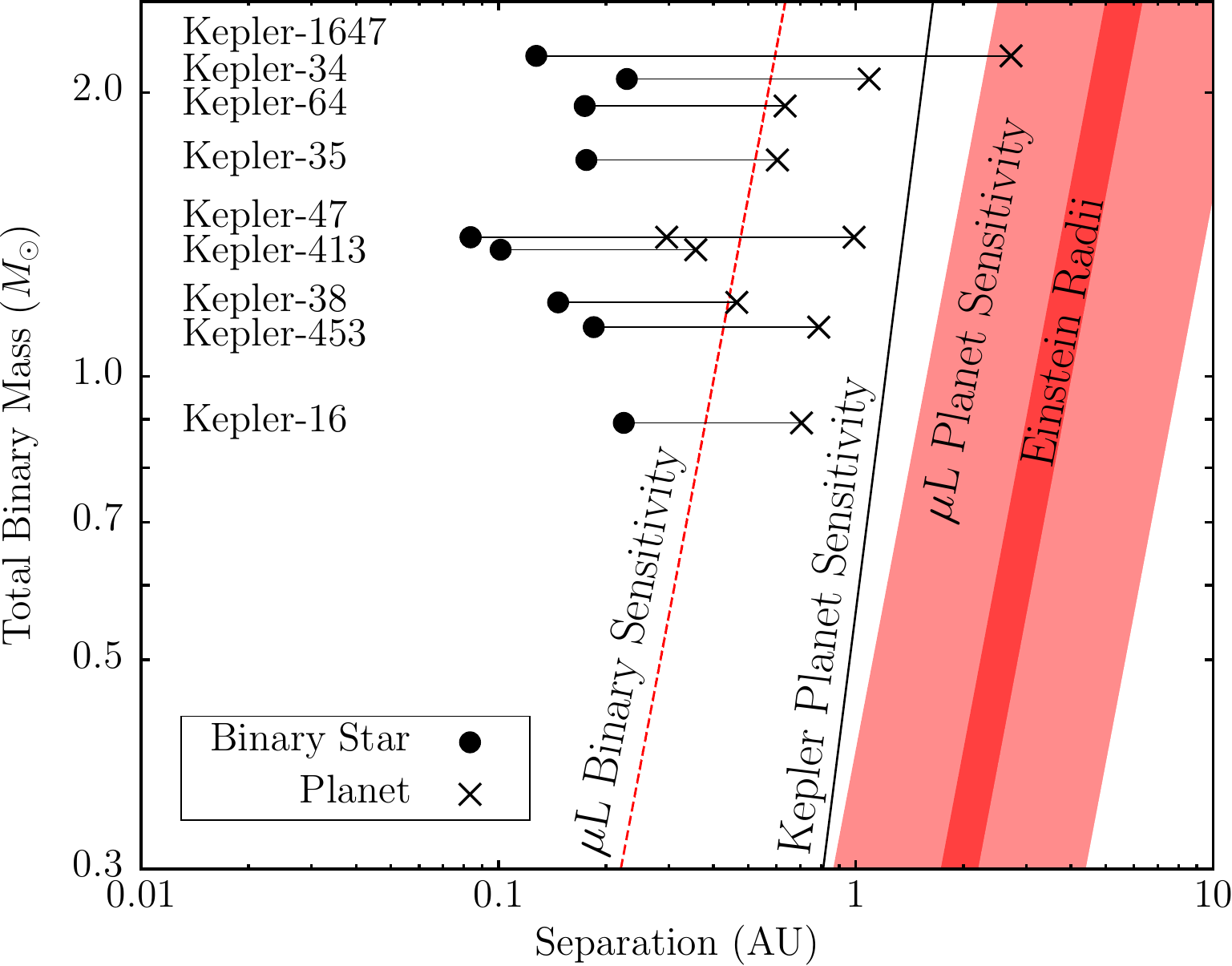} 
\caption[Separations of known circumbinary planets from \kepler\ data.]
{\singlespace  Stellar binary and circumbinary planet separations of the circumbinary systems found using \kepler\ data. The large dots plot the separation between the two stars in the stellar binary and the crosses show the separation of the circumbinary planet(s) from the center of mass of the binary for each binary system. Each binary star is linked to its planet(s) with a line. The dark red region shows the range of Einstein radii for typical lens and source distances when viewing toward the Bulge of the Milky Way. The light red region shows the approximate range of microlensing's sensitivity (i.e. 0.5 $\to$ 2 $r_{\mathrm{E}}$). The red dashed line shows the approximate lower limit of microlensing's sensitivity to binary stars (${\sim}0.1r_{\mathrm{E}}$). The solid black line shows the approximate limit of \emph{Kepler}'s sensitivity to circumbinary planets, for which we assume the planet must complete at least three orbits during the $\sim$4 year \kepler\ mission duration. Note that Kepler-1647b was detected having completed less than 2 orbits. Circumbinary planet data has been taken from \citep{Doyle2011,Welsh2012,Orosz2012,Orosz2012a,Schwamb2013,Kostov2014,Welsh2015, Kostov2015}.}
\label{kepler}
\end{figure}

To compare the range of sensitivity to circumbinary planets between transit and microlensing surveys, in \autoref{kepler} we plot the projected separations and total binary mass of the nine circumbinary planetary systems found by \kepler\ and compare them to the range of microlensing's sensitivity to planets. Specifically, the dark red band shows the interquartile range of $\re$ from the simulations of \citep{Henderson2014-kmt}, ($3.2$--$4.0$)$(M/\msun)^{-1/2}$, and the lighter red band shows this range expanded by a factor of $2$ in each direction to indicate approximate region of planet sensitivity~\citep[i.e., slightly wider than the traditionally defined lensing zone][]{Wambsganss1997, Han2009}. We take the range of sensitivity to binaries to be $>0.1\re$. Because \kepler\ targetted largely solar-mass stars, the \kepler\ circumbinaries lie in the top part of the plot. However, because microlensing is sensitive to planets around stars of any mass, it should be sensitive to circumbinary planets in binaries of lower mass as well~\citep[e.g., M-M binaries][]{Shan2015}. Additionally, the \kepler\ circumbinaries have orbits that are slightly too small to be detectable via microlensing, but planets on slightly wider orbits around slightly wider binaries should be detectable. Planets on wider orbits around close binaries with binary separations similar to those seen in the \kepler\ population of circumbinaries would also be detectable via microlensing, but it is unlikely that the binary nature of the host would be recognizable, unless the hosts were monitored with radial velocity observations, which given the masses of at least some of them should be possible~\citep[e.g.,][]{Boisse2015, Yee2016-rv}. Transits (and thus \kepler) have a strong bias toward detecting planets with short period orbits, both because the transit probability is higher, and because there are more transits.  Therefore, most circumbinary planets found by \kepler\ are found at the innermost stable orbit \citep[see, e.g.,][]{Winn2015,Holman1999}. Further, \kepler\ is completely insensitive to planets with perods $> \sim$2 years, because it requires at least two transits. 

 The lightcurve of a microlensing event largely depends on the structure of the caustics. The caustics are closed, cuspy curves in the source plane that demark regions with differing numbers of lensed images; they are a tell-tale sign of a system with more than one lens (or of a lens with external shear, although often this shear is due to another point lens). Because the images are not resolved, the observational consequence of caustics in a microlensing event is a pair of sharp peaks in the lightcurve when the source passes over them. The caustics themselves are the set of points in the source plane where the magnification is formally infinite, but because source stars are not point like, the magnification can be extreme (${\sim}1000$ or more) but remains finite. The caustics for an $N$-body lens can be found by solving the equation $\det \mathbf{J} = 0$, where $\mathbf{J}$ is the Jacobian of the lens equation, which represents the many to one mapping of images to source positions. The lens equation for a lens with $N$ point masses can be written as
\begin{equation}
\zeta = z - \sum_{j}^{N}{\frac{m_j}{\overline{z} - \overline{z}_{j}}},
\label{lensequation}
\end{equation}
where we have used the standard complex formalism~\citep{Witt1990} and units of Einstein radius of a mass $M$, with $\zeta$ being the location of the source, $z$ the location of an image, $z_j$ and $m_j$ the location and nomalized mass (relative to $M$) of each lens $j$; $N$ is the number of lenses and the overline represents complex conjugation. $\det \mathbf{J}=0$ can be rearranged to
\begin{equation}
\sum_j^N{\frac{m_j}{(\overline{z} - \overline{z}_j)^2}} = e^{i\psi},
\label{criticalcurvesoln}
\end{equation}
a $2N^{\text{th}}$ order polynomial with a parameter $\psi$ that runs from $0$ to $2\pi$. The solutions to this equation when run over the range of $\psi$ form a set of curves in the image plane known as critical curves, which are the positions where pairs of images can be created or destroyed. The caustics are found by mapping the critical curves back to the source plane through the lens equation. 

The simplest caustics are double-lens caustics (e.g., a star plus planet, or a binary star), which, with only two parameters, have been well studied theoretically~\citep[e.g.,][to name but a few]{Schneider1986,Bozza1999,Dominik1999,Han2006}. Circumbinary planet systems consist of a planet orbiting around a binary star and are therefore triple-lens systems with more complicated caustics. 

The parameter space of triple-lens systems is significantly larger with five parameters, and while there have been several studies, the exploration of the full parameter space is far from complete. \citet{Rhie2002} wrote down the lens equation for a triple-mass lens, and there have been several theoretical studies of $N$-body lenses, either where symmetries make the system analytical tractable~\citep[e.g.,][]{Mao1997, Rhie2003, Asada2009, Danek2015-triple} or using perturbative analysis~\citep[e.g.,][]{Bozza2000-special,Bozza2000-secondary,Asada2009a}, and in some cases more general studies~\citep[e.g.,][]{Danek2015-general}. The triple-lens configuration receiving the most attention has been multiplanet systems around single stars~\citep[e.g.,][]{Gaudi1998,Han2001,Ryu2011,Song2014,Zhu2014-parameters,Zhu2014-fitting}, and two examples of these have been discovered~\citep{Gaudi2008, Bennett2010, Han2013}. Star-planet-moon systems have also received some attention~\citep{Han2002, Han2008-moon, Liebig2010}, though no examples have been discovered yet. Binary star systems with planets have received less theoretical attention. Circumprimary systems have been examined by \citet{Lee2008}, but notably \citet{Poleski2014} and \citet{Gould2014} have recently found examples of such planets. \footnote{Microlensing event OGLE-2013-BLG-0723 \citet{Udalski2015-venus} was claimed to be caused by a circumprimary planet, but it has since been shown to be consistent with just a binary star undergoing orbital motion \citep{Han2016}. Additionally, \citet{Penny2016-devoid} argued that the distance to the lens implied by the circumprimary model of OGLE-2013-BLG-0723 was much smaller than one would typically expect.}

 To date, only one microlensing event has been claimed to be caused by a circumbinary planet, MACHO-97-BLG-41L \citep{Bennett1999}, but this was later shown to be caused by binary stars with orbital motion \citep{Albrow2000,Jung2013,Han2016}. There have been no other microlensing circumbinary planet detections published, and relatively little theoretical work on microlensing by circumbinary planets, with only a brief paper by \citet{Han2008-tatooine} addressing them directly. By assuming that circumbinary planets would be most detectable when the size of the central caustics of the binary and planet were similar, \citet{Han2008-tatooine} estimated the range of planet and binary separations for which the microlensing technique is efficient at detecting circumbinary planetary systems. He found that stellar binaries with semimajor axis $a_{\rm b}$ between $0.15$ and $0.5$~AU and Jupiter-mass planets with semimajor axis $a_{\rm p}$ between $1$ and $5$~AU were optimal, with smaller ranges for lower mass planets.  

In this work we make a number of inroads into the topic of circumbinary planet microlensing. In \autoref{caustics} we begin by exploring the phenomenology of the caustics in the circumbinary system, culminating in an analytic expression for the location of planetary caustics. In \autoref{orbitalmotion} we consider the impact of the binary's orbital motion relative to the planet's on caustic motion. In \autoref{fractionaldetectability} we build upon the \citet{Han2008-tatooine} study and attempt to quantify the fraction of circumbinary planet events with caustic crossings that will be obviously recognizable as triple lenses. We end the paper with a brief discussion in \autoref{discussion} and conclusions in \autoref{conclusions}.

\section{Caustics of Circumbinary Microlenses}\label{caustics}

In this section we aim to gain an understanding of circumbinary caustics, first by highlighting unique features of circumbinary caustics (relative to both planetary and binary microlenses) and then by developing more quantitaive descriptions. Understanding the structure of circumbinary caustics is important for understanding potential circumbinary microlensing events. The circumbinary parameter space is larger than for double lenses, and so we find that the most efficient way to survey it and gain an intuitive understanding of the caustics is by creating a series of animations scanning through various parameters. Before we begin though, a primer on double-lens caustics is useful to set the stage and it will be necessary to define our parameterization of the circumbinary system. 

\subsection{Primer on Notation and Double-Lens Caustics}\label{primer}

In the context of this paper, the word binary can be confusing, e.g., when used in the term ``binary lens'' it can refer to any two-lens system including a planet and a star, but when used in the term ``binary star'' it is more restrictive. In order to avoid confusion, we will refer to the three distinct lens systems with the following notation:
\begin{itemize}
\item ABb -- a circumbinary planet system, composed of two stars and a planet,
\item AB -- a binary star, composed of two stars,
\item Ab -- a system composed of a planet and a star.
\end{itemize}

\begin{figure}
\includegraphics[width=0.5\textwidth]{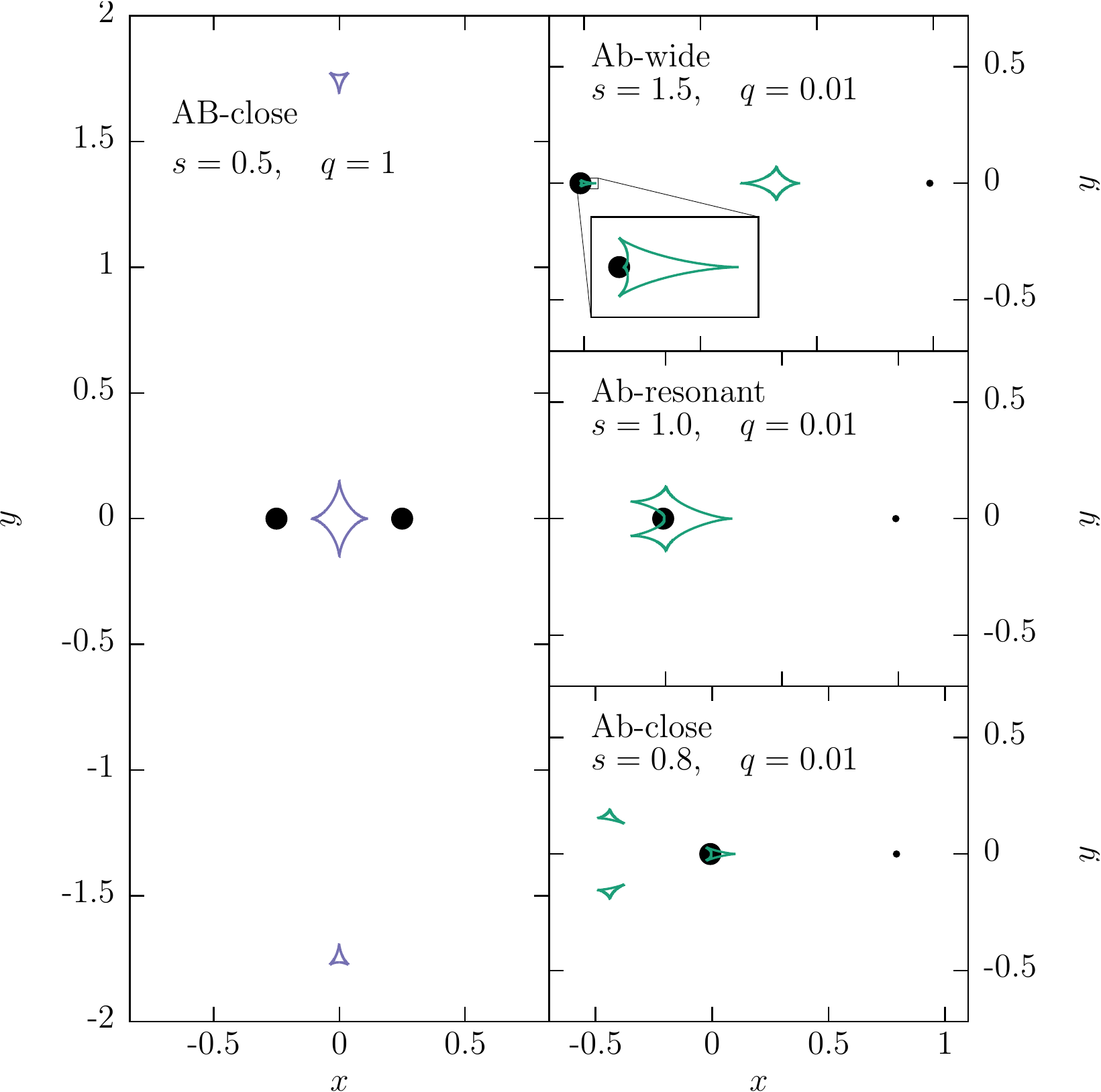}
\caption{Caustics of the double lens. The left panel shows the close topology caustics of a binary star ($q\sim 1$). The right panels show the close resonant and wide topologies of planetary lenses (i.e., planet and a star, $q \ll 1$). The inset shows the central caustic of the wide planetary lens, which is nearly identical to the central caustic of the close planetary lens with separation $1/s$. The circles show the positions of the lenses. Units are normalized to the Einstein ring.}
\label{doublecaustics}
\end{figure}

A double-lens model (e.g., a single star with a planet, or a binary star) can be fully described by two parameters: the mass ratio of the two bodies, $q$, and their projected separation in units of the Einstein ring, $s$. The caustics of double lenses can have one of three topologies, named \textit{close}, \textit{resonant}, and \textit{wide}~\citep{Schneider1986, Erdl1993}. For small mass ratios, $q \ll 1$, the boundaries between the close and resonant topologies lies at $s \simeq 1-3q^{1/3}/4$ and the boundary between resonant and wide is at $s \simeq 1+3q^{1/3}/2$; resonant topologies have $s \sim 1$~\citep{Dominik1999}. For equal-mass bodies, $q=1$ and the boundaries lie at $s = 1/\sqrt{2}$ and $s = 2$~\citep{Schneider1986}. \autoref{doublecaustics} shows examples of each of the double lens caustics we will encounter in this work: that of a close, binary lens (AB, $q\sim 1$) and close, resonant and wide planetary lenses (Ab, $q\ll 1$). Close lenses have three caustics: one \emph{central} caustic near the center of mass, and two \emph{planetary} (in the Ab system) or \emph{secondary} (in the AB system) caustics. Resonant lenses have just one large caustic near the center of mass, and wide lenses have a central caustic very similar to the central caustic of the close lens and a single planetary caustic. For the most part, the ABb systems studied here have the same close, resonant, and wide characterizations that depend on the separation of the circumbinary planet, and so we will use this same terminology to loosely describe the ABb topologies; for a more exact definition of the caustic topologies of multi-body lenses, see \citet{Danek2015-general, Danek2015-triple}.

\subsection{Circumbinary Parameter Space}\label{parameterspace}

\begin{figure}
\includegraphics[width=\columnwidth]{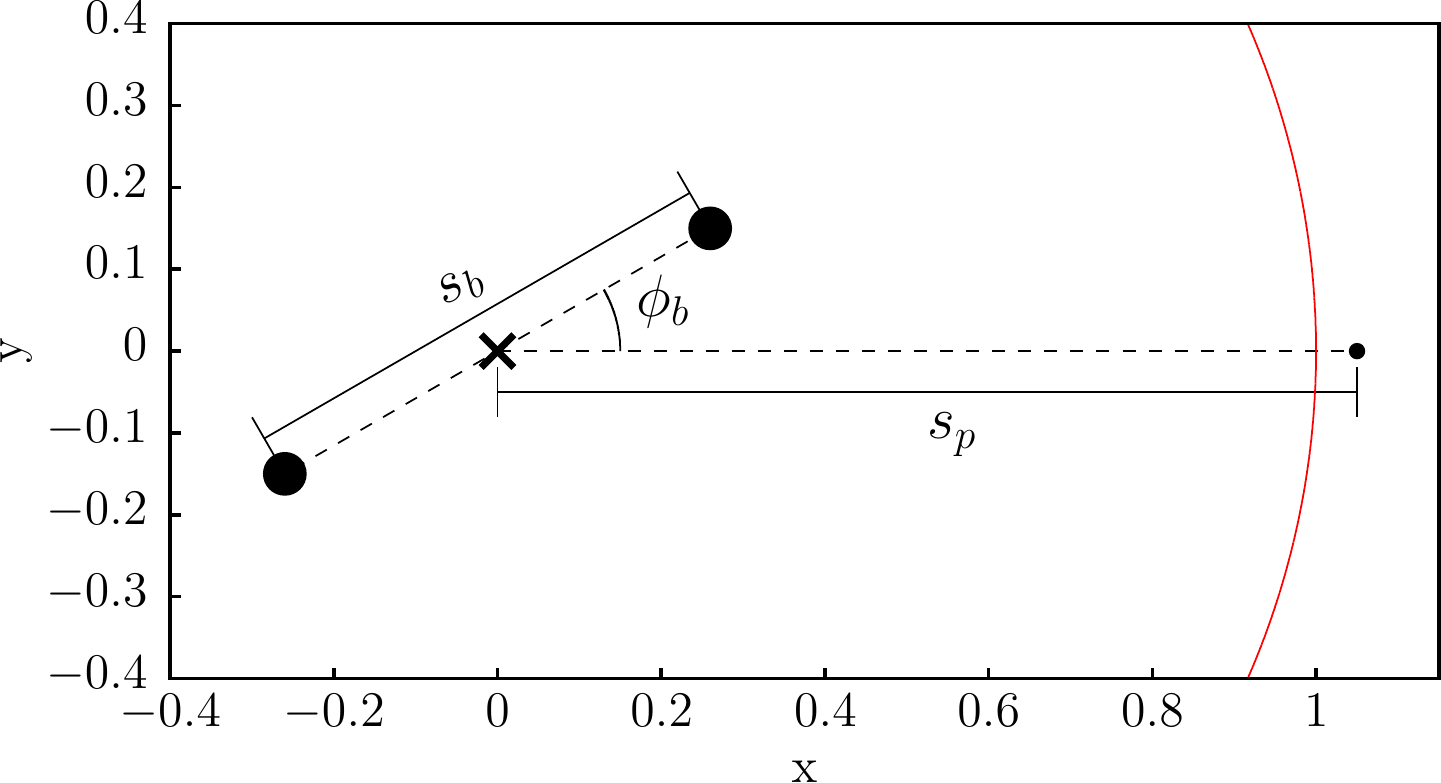} 
\caption[Diagram showing the coordinate system and parameters used for calculations]
{\singlespace Parameters and reference frame of the circumbinary lens. The circles mark the locations of the lenses, with the planet being the smaller of the three. The thick X marks the center of mass of the stellar binary, which we define to be the origin of our coordinate system.  Note that this is not the center of mass of the whole 3-body system. The red circle indicates the Einstein ring.}
\label{diagram}
\end{figure}

 With triple-lens circumbinary planet systems, the additional lens increases the number of necessary parameters to five; we have chosen these to be the projected separations of the binary star $s_{\mathrm{b}}$ and planet $s_{\rm p}$, in units of the Einstein radius, the mass ratios of the binary star $q_{\mathrm{b}} \equiv m_{\rm A}/m_{\rm B}$ and planet $q_{\rm p}  \equiv m_{\rm p}/(m_{\rm A} + m_{\rm B})$, and the angle of the line connecting the two binary components to the line connecting the planet to the binary center of mass $\phi_{\rm b}$.

We work in a coordinate system with the origin located at the \emph{center of mass of the binary} and with the circumbinary planet located on the x-axis at a distance $s_{\mathrm{p}}$. This means that $\phi_{\mathrm{b}}$ is simply the angle of the binary from the x-axis. \autoref{diagram} shows a diagram of the system and the parameters. We define the Einstein radius to be that of the combined mass of the binary star \emph{excluding the planet}. These choices were made to ensure there would be no shift or rescaling needed when comparing the ABb to AB and Ab systems.

\subsection{Animating Caustics as a Function of $s_{\rm b}$ and $\phi_{\rm b}$}\label{animations}

The large number of parameters of the ABb system makes it difficult to adequately explore the entire parameter space. For that reason, we made a collection of animations to quickly gain a more intuitive understanding of how the caustics change as the projected separation and angle of the binary star change. We have created a grid of 60 animations, half of which use $s_{\rm b}$ (varying between $0$ and $0.4$) as the time variable and the other half use $\phi_{\rm b}$ (running from $0$ to $2\pi$). For each time variable, we explore a $5\times 3 \times 2$ grid covering the other three parameters of the ABb system, $s_{\rm p}$, $q_{\rm p}$ and $q_{\rm b}$, respectively. The values of $s_{\rm p} =  0.6, 0.95, 1.0, 1.05$ and $1.5$ sample close, resonant and wide topologies, and the values of $q_{\rm p}=10^{-3}, 10^{-4}, 10^{-5}$ sample a range of planet masses corresponding roughly to Saturn, Neptune and Earth-mass planets orbiting a binary with total mass of $0.3\msun$. $q_{\rm b}$ takes values of $1.0$ and $0.3$. The entire collection of animations can be viewed at \url{http://www.astronomy.ohio-state.edu/~luhn.5/animations/}.

In addition to the circumbinary caustics, the animations also show the caustics of the AB and Ab systems for comparison. The AB system is just the ABb system with the planet removed. The Ab system is a planet-star double lens, but the star has a mass equal to that of the combined binary mass in the ABb system. \autoref{diagram} can be used to show the three systems that are compared here: ABb, AB, and Ab. The three points represent the full ABb system. The AB system can be seen by taking away the planet (smaller point), leaving the two larger points. The Ab system can be seen by taking the two large points and replacing them with a point at the \textbf{X}. 

The two double-lens systems are important to understanding the full triple-lens circumbinary system because the caustics produced from the ABb system are often well-approximated by a superposition of the AB and Ab systems \citep{Han2008-tatooine}, and will tend towards one or the other if one of the various separation and mass ratio parameters approach zero. By comparing the ABb caustics to the AB and Ab caustics it is possible to get a qualitative idea of whether the triple-lens circumbinary system would be recognizable as such in the lightcurve of an event (which can be thought of as a 1-dimensional slice through the caustics) or if it could be mistaken for one of these double-lens systems.

\subsubsection{Understanding the Animations}

\begin{figure*}
\includegraphics[width=\textwidth]{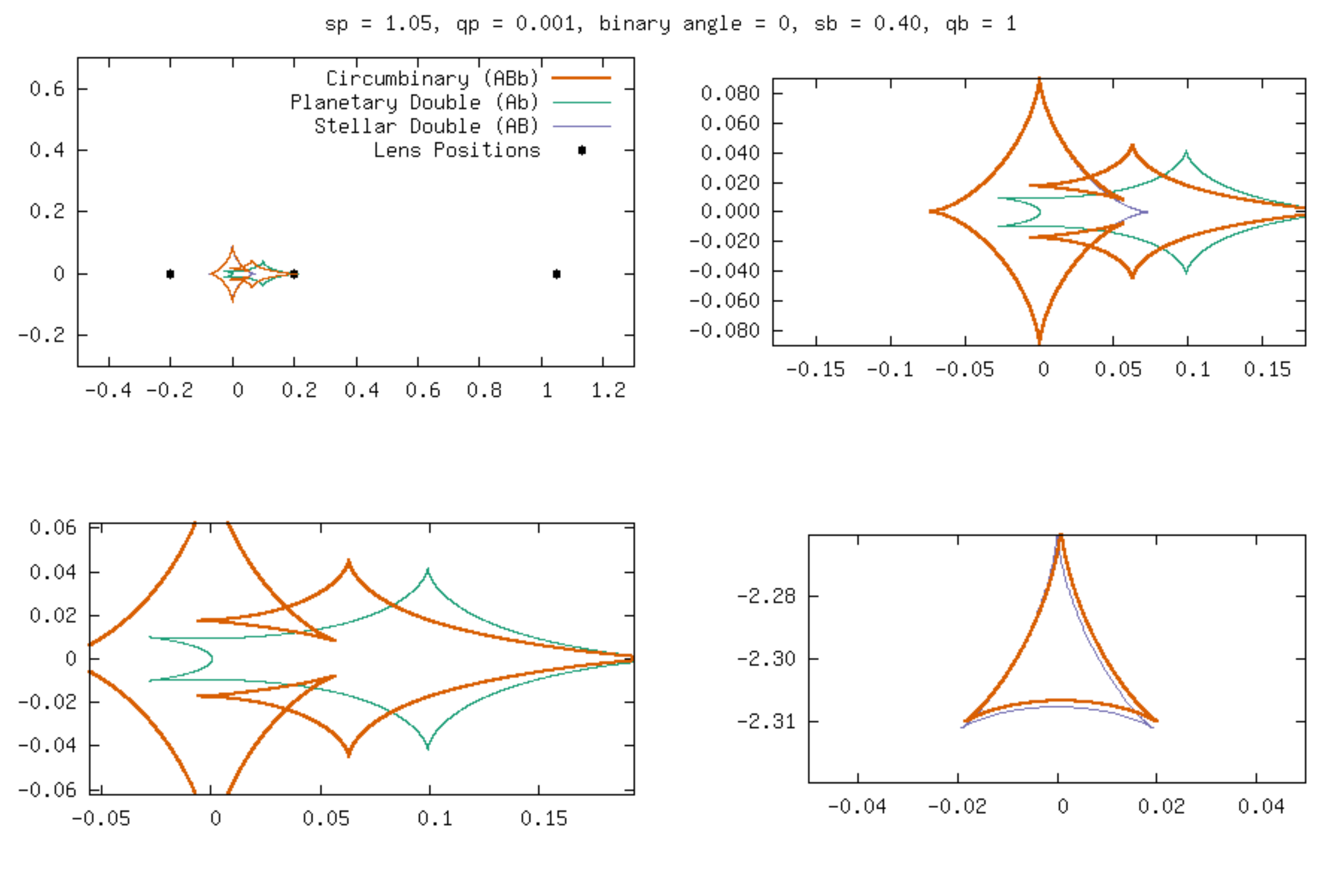} 
\caption{Screenshot of one of our animations comparing the circumbinary (ABb, orange) caustics to those of the stellar binary (AB, purple) and the planetary double lens (Ab, green). For further distinction between circumbinary caustics and the two double lens approximations, the circumbinary ABb caustics are plotted with a thick line. The top left panel shows the whole system, except for two of the stellar binary caustics which are located above and below the window (which can be seen in the left panel of \autoref{doublecaustics}). The top right panel zooms in on the central caustic. The bottom left panel focuses on the planetary caustics (and thus only compares ABb and Ab caustics). The bottom right panel zooms in on one of the two stellar binary caustics which were not shown in the top left panel. All of the animations can be found at \url{http://www.astronomy.ohio-state.edu/~luhn.5/animations/}.}
\label{lenscomparison}
\end{figure*}

\autoref{lenscomparison} shows a sample screenshot of an animation. Each animation has four panels. The \emph{top left} panel gives an overview of the all of the caustics except the two small secondary caustics of the stellar binary that lie far \citep[${\sim}1/s_{\mathrm{b}}$][]{Bozza2000-secondary} from the center of mass. The window size and location (pan and zoom) of the top left panel is constant throughout the animation, but may vary slightly over the grid. The \emph{top right} panel shows the central caustic region, usually comparing the ABb caustic to the AB caustic. In the $\phi_{\mathrm{b}}$ animations, the window stays fixed. However, in the $s_{\mathrm{b}}$ animations, the window zooms in and out to keep the size of the central caustic of the stellar binary roughly fixed relative to the window size, but will abandon this once it becomes too small. Occasionally this window will pan. The \emph{bottom left} panel focuses on the planetary caustics. In the cases where there are two planetary caustics, this panel might only focus on one of them, since the other moves in a similar way. For the $\phi_{\mathrm{b}}$ animations, this window stays in a fixed position. For the $s_{\mathrm{b}}$ animations, the window stays a fixed size but pans to keep the ABb planetary caustic in the center of the window (it might jump if there is a topology change). This is done to highlight changes in the size of the caustics. The \emph{bottom right} panel focuses on the lower of the two far triangular caustics from the stellar binary that are not shown in the first panel. In both $s_{\mathrm{b}}$ and $\phi_{\mathrm{b}}$ animations, the window stays a fixed size, but moves so that the AB caustic remains in the center of the window. In each panel, it is important to take note of the axes and how they are changing in order to understand what each window is doing.

The animations provide an efficient way to explore the circumbinary parameter space. What follows is a description of the interesting features we observed from the animations.

\subsection{Superposition, Self-Intersection and Metamorphoses}\label{superposition}

What is immediately noticeable from the animations is the fact that often the ABb system is well-approximated by a superposition of the AB and Ab systems, at least in parts of the caustic, as has been previously noted by \citet{Han2008-tatooine} and similarly noted for multiplanet systems by \citet{Gaudi1998} and \citet{Han2001}. In fact, the ABb caustics appear to largely follow the three topologies of double-lens planetary microlensing, though see \citet{Danek2015-triple} for a more precise definition of the multiple possible topologies of triple lenses.

The superposition is not always an exact superposition, or might be better described by the superposition of lenses that are at different relative positions or have slightly different parameters. In these cases, superposition is most helpful as a good first assumption. In the cases where the AB and Ab systems have caustics in the same location, the ABb caustic is dominated by the system with the larger caustic (AB or Ab). However, in situations where the AB and Ab caustics are approximately the same size, the resulting ABb caustics can self-intersect and undergo swallowtail metamorphoses \citep{Schneider1992,Petters2001,Danek2015-general}. Butterfly metamorphoses are also possible (the $\phi_{\rm b}$ animation with $s_{\rm p}=0.95$ and $q_{\rm p}=10^{-4}$ shows a nice example when $\phi_{\rm b}=90\degr$ and can be found at \url{http://www.astronomy.ohio-state.edu/~luhn.5/animations/sp_0.95_qp_0.0001_phi.gif}), but are much less common than swallowtails, so we will focus on the swallowtails. We will also use the terms swallowtail and butterfly more loosely than their strict definition as catastrophes, often using them to describe regions where caustic self intersection causes regions of with at least four more images than the minimum number of images. It is at the locations of these swallowtail and butterfly regions caused by self intersection that the triple-lens' nature is most easily recognizable as nested caustic crossings (i.e., a pair of caustic crossings inside another caustic crossing pair), because the caustics of an isolated two point mass lens system cannot intersect \citep{Schneider1986}.

\begin{figure}
\includegraphics[width=\columnwidth]{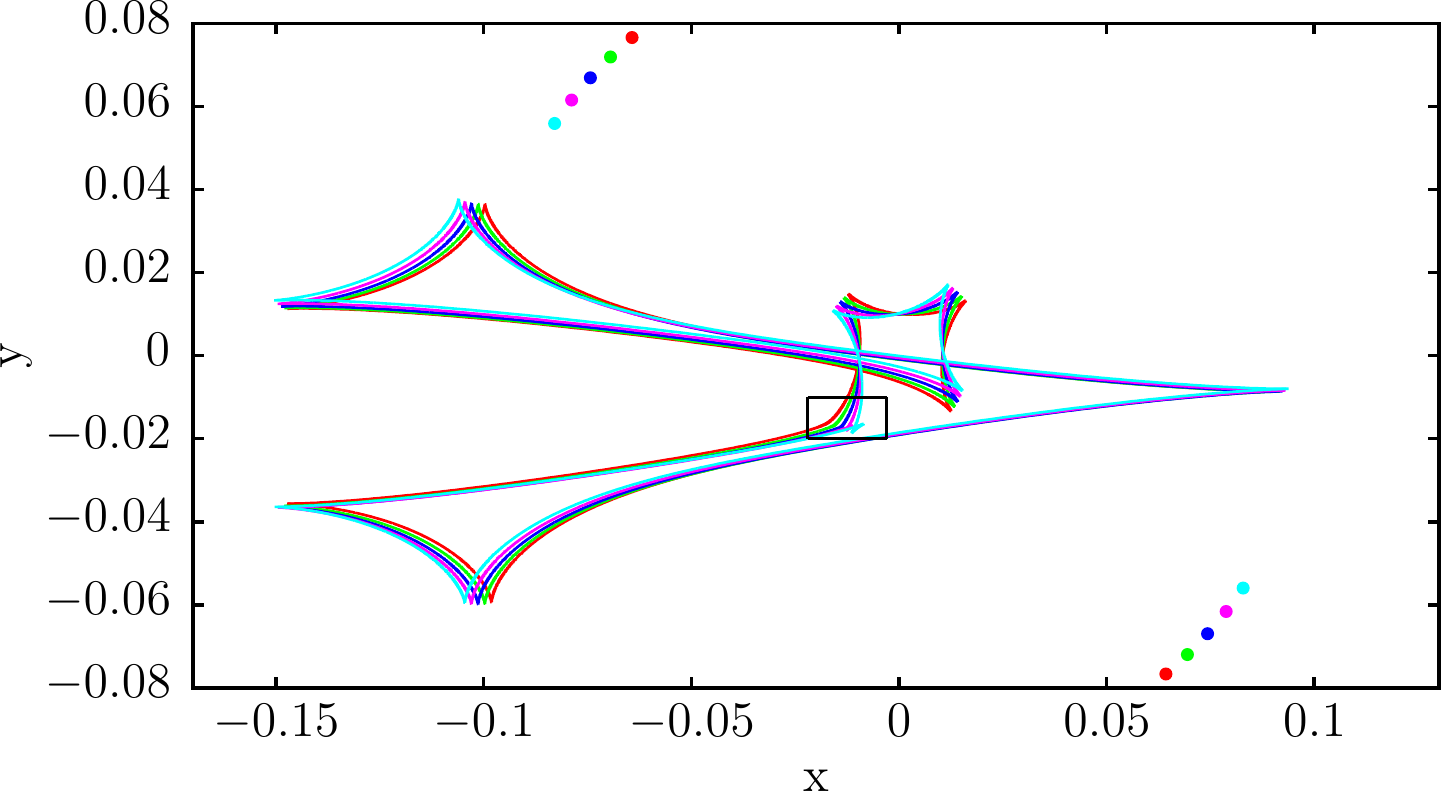}
\\
\\
\includegraphics[width=\columnwidth]{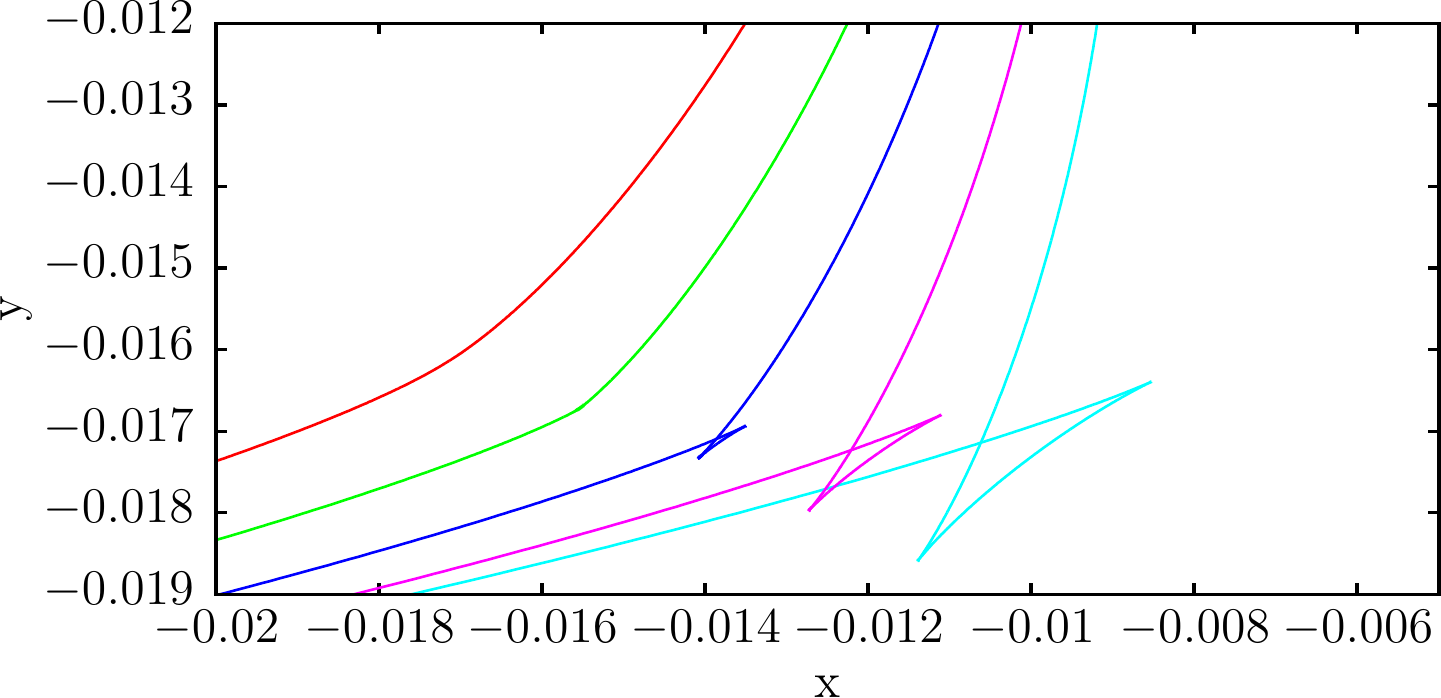}
\caption{The progression of a swallowtail metamorphosis in the roughly resonant ABb caustic as $\phi_{\mathrm{b}}$ changes. The top panel shows the full central caustic while the bottom panel zooms in on the boxed region to show creation of a swallowtail. Gradually the swallowtail grows to resemble part of the AB caustic. The colored points on the top panel show the position of the stellar binary lenses as $\phi_{\mathrm{b}}$ changes from $130^{\circ}$ (red) to $146^{\circ}$ (cyan). The colors of the points correspond to the stellar binary positions of the same-colored caustic. Here $s_{\mathrm{b}}$ = 0.2, $s_{\mathrm{p}}$ = 0.95, $q_{\mathrm{b}}$ = 1, and $q_{\mathrm{p}}$ = 10\textsuperscript{-3}. The corresponding animation can be seen at \url{http://www.astronomy.ohio-state.edu/~luhn.5/animations/sp_0.95_qp_0.001_phi.gif}.}
\label{cusp_creation}
\end{figure}

\autoref{cusp_creation} shows the creation of a swallowtail as $\phi_{\rm b}$ changes. We find that the swallowtail metamorphosis occurs only on the resonant and central caustics of the ABb system, but not on the planetary caustics if they are separated from the central caustic. The swallowtails and butterflies can be as large as the caustic on which they reside, when the AB and Ab central caustics have similar sizes. The actual point at which the metamorphosis occurs (e.g., when a swallowtail is created or destroyed) tends to be where a fold of the ABb caustic lies close to position of one of the cusps of the AB central caustic that point towards its secondary caustics. The AB central caustic cusps that point towards the binary lens components do not produce metamorphoses.

\begin{table*}
\caption{Properties of binary caustics and their limiting behavior}
\label{tbl:topologies}
\centering
\begin{tabular}{C{15em}C{10em}C{10em}C{10em}cc@{}}
\hline\hline
Topology 			& Close			& Resonant 			& Wide & Reference &\\ [3ex]
Number of Caustics 		& 3			& 1 				& 2    & \citet{Schneider1986}  &\\ [3ex]
Topology Boundary 		& $s \simeq 1- \frac{3q^{1/3}}{4}$ & $s_{\mathrm{close}} < s < s_{\mathrm{wide}}$		& $s \simeq 1+ \frac{3q^{1/3}}{2} $ & \citet{Dominik1999} &\\ [3ex]
Central Caustic Shape 		& Arrowhead		& Merged with planetary & Arrowhead & ------ &\\ [3ex]
Central Caustic Number of Cusps & 4 			& 6 (merged caustic) 		& 4 & ------ &\\ [3ex]
Central Caustic Size 		& $\propto s^1 q$	& ------			& $\propto s^{-1} q$ & \citet{Griest1998a} &\\ [3ex]
Planetary Caustic Shape		& Triangular		& Merged with central	& Astroid & ------  &\\ [3ex]
Planetary Caustic Number of Cusps & 3			& 6 (merged caustic)		& 4 & ------  &\\[3ex]
Planetary Caustic Size		& $\propto s^3 q^{1/2}$	& ------			& $\propto s^{-2}q^{1/2}$ & \citet{Han2006}  &\\[3ex]
Planetary Caustic Position	& $ \left(\frac{1-q}{1+q}\right)\left(s-1/s\right)$ & ------	& $ \left(\frac{1-q}{1+q}\right)\left(s-1/s\right)$ &  \citet{Bozza2000-special}  &\\[3ex]
\hline
\end{tabular}
\end{table*}

\subsection{Central Caustics of Close and Wide Lenses}\label{centralcaustics}

Central caustics appear in all three topologies: close, resonant, and wide, but are largest in the resonant configuration. Refer to \autoref{tbl:topologies} for a review of the properties of the caustic topologies and their limiting behaviors. When in the close and wide topologies, the central caustics of the Ab system are four-cusped, arrowhead-shaped caustics, unlike the central resonant caustic, which has six cusps. Although the AB caustics are still astroid-shaped with four cusps, the superposition of the astroid with the arrowhead behaves differently than the superposition of the six-cusped resonant shape with the astroid. 

\begin{figure*}
\includegraphics[width=\columnwidth]{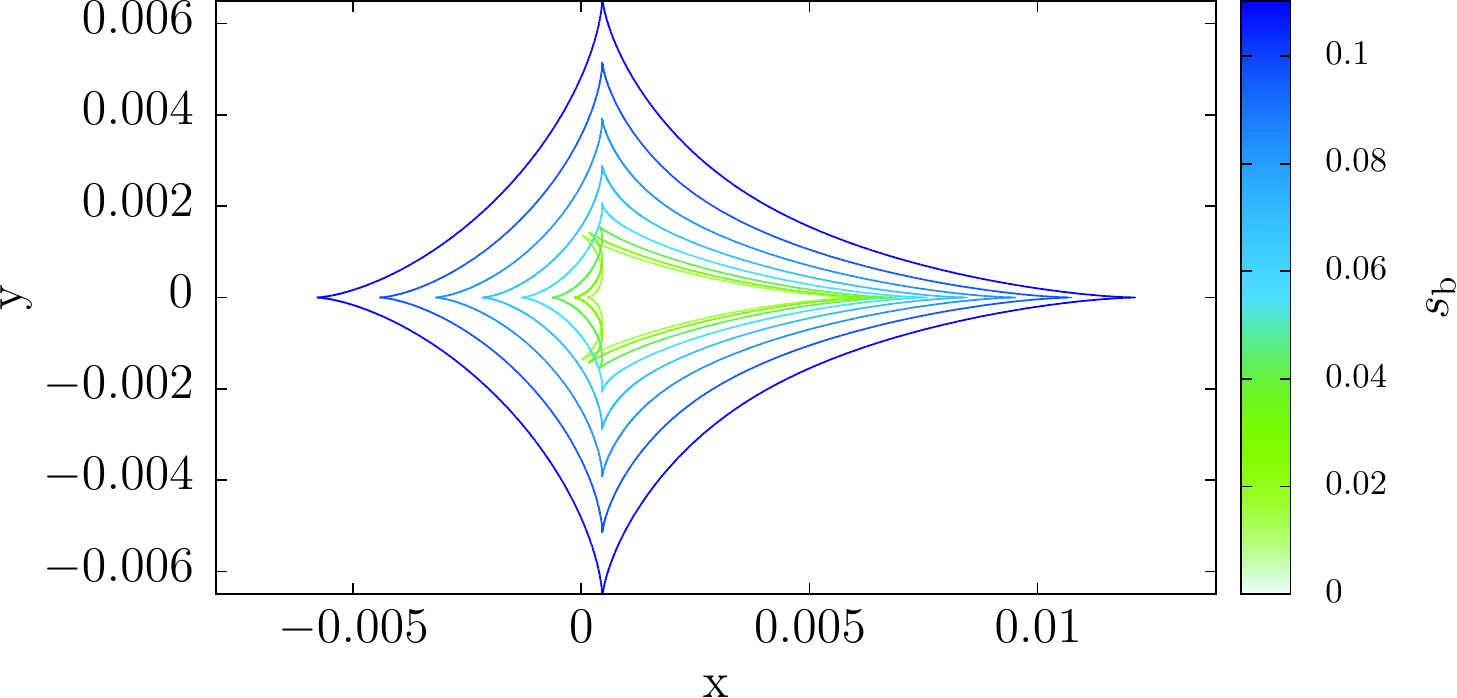}
\hspace{14pt}
\includegraphics[width=\columnwidth]{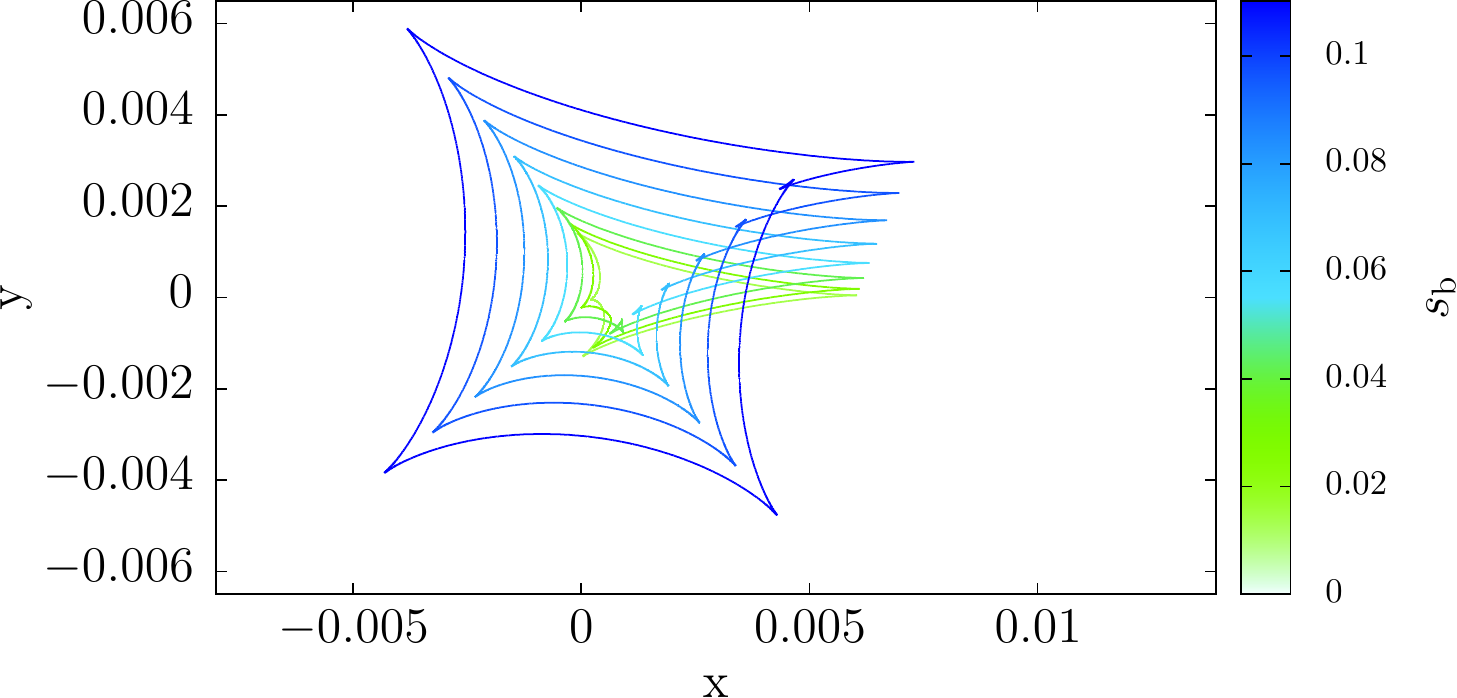}
\\
\vspace{12pt}
\\
\includegraphics[width=\columnwidth]{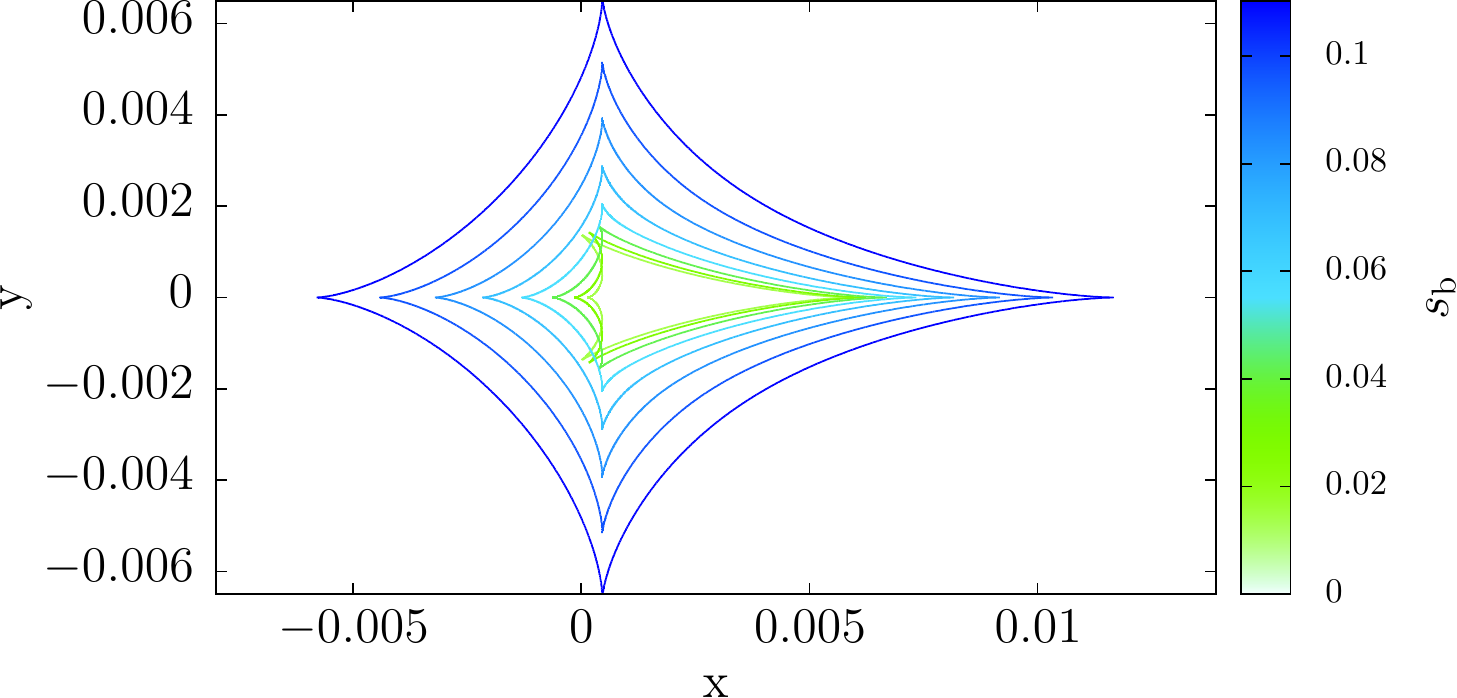}
\hspace{13pt}
\includegraphics[width=\columnwidth]{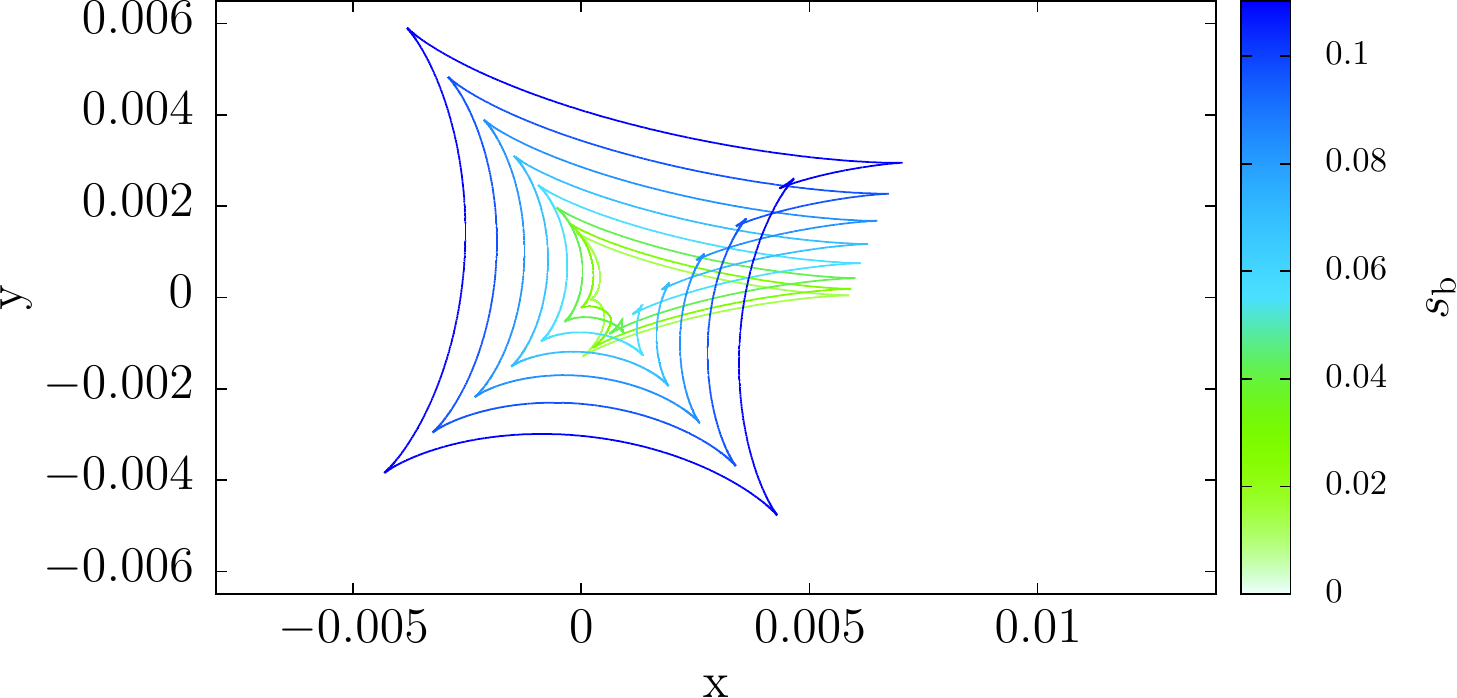}
\caption{Central caustics of the close and wide topologies as $s_{\mathrm{b}}$ varies, with $s_{\mathrm{b}}$ encoded by color. As the color changes from blue to green, the caustic shapes changes from more astroid-shaped to more arrowhead-shaped as it moves from AB to Ab dominated. The ABb caustic shape is intermediate to those from the AB and the Ab systems since they are neither fully arrowhead-shaped nor fully astroid-shaped. The left panels show the central caustics for $\phi_{\mathrm{b}}=0^{\circ}$ and the right panels show the central caustics for $\phi_{\mathrm{b}} = 40^{\circ}$. The top panels show the central caustic of a wide topology ($s_{\mathrm{p}}$ = 1.5), and the bottom panels show the central caustic of a close topology ($s_{\mathrm{p}}$ = 1/1.5 = 0.67). Notice that these two topologies are not distinguishable from each other because of the degeneracy between $s_{\mathrm{p}}$ and $1/s_{\mathrm{p}}$ for the close and wide topologies. These plots use $q_{\mathrm{b}}=1$ and $q_{\mathrm{p}}=10^{-3}$.}
\label{central_caustic}
\end{figure*}

Large swallowtails occur less frequently in the close and wide topologies. This is likely due to the fact that the central caustics in these regimes are significantly smaller than the resonant caustic. Additionally, unlike in the resonant caustic where the regions enclosed by self intersection can change size significantly, these will often remain of a similar size with changing $\phi_{\rm b}$ and even surprisingly $s_{\rm b}$. This can be seen in \autoref{central_caustic}, and in the closest and widest $\phi_{\rm b}$ animations, where a distortion to the central caustic is centered on a small swallowtail (unresolved in the animations) that moves around the caustics as $\phi_{\rm b}$ changes. In addition swallowtails are not always present on the central caustic, e.g., for angles of $\phi_{\mathrm{b}}$ close to $0^{\circ}$. Despite this, the ABb central caustic can still differ significantly from the AB and Ab central caustics. However, the familiar $s\leftrightarrow 1/s$ degeneracy~\citep{Dominik1999, An2005} still afflicts circumbinary planets. As can be seen in the figure, the circumbinary caustics with $s_{\rm p}$ and $1/s_{\rm p}$ are essentially identical \citep{Griest1998}.

For angles of $\phi_{\mathrm{b}}$ close to $0^{\circ}$, the shape of the ABb caustic is interesting. For these small angles, the caustic has only four cusps and does not self-intersect, which gives it a resemblance to the central caustics of a wide, equal mass binary: asymmetrically elongated along the $x$-axis, but with the off-axis cusps pointed vertically instead of slightly towards negative $x$-values.  The overall shape of the ABb caustic is determined by which of the two double-lens central caustics is dominant: when the central caustic of AB is larger than the central caustic of Ab, the shape of the central caustic looks more like the central caustic of AB, and vice versa. \autoref{central_caustic} shows the progression of the ABb central caustics from AB dominated (bluer, more astroid-shaped) to Ab dominated (greener, more arrowhead-shaped). It is unlikely that the lightcurve of a source trajectory crossing this caustic will be immediately recognizable as a triple system without detailed modelling.

\begin{figure}
\includegraphics[width=\columnwidth]{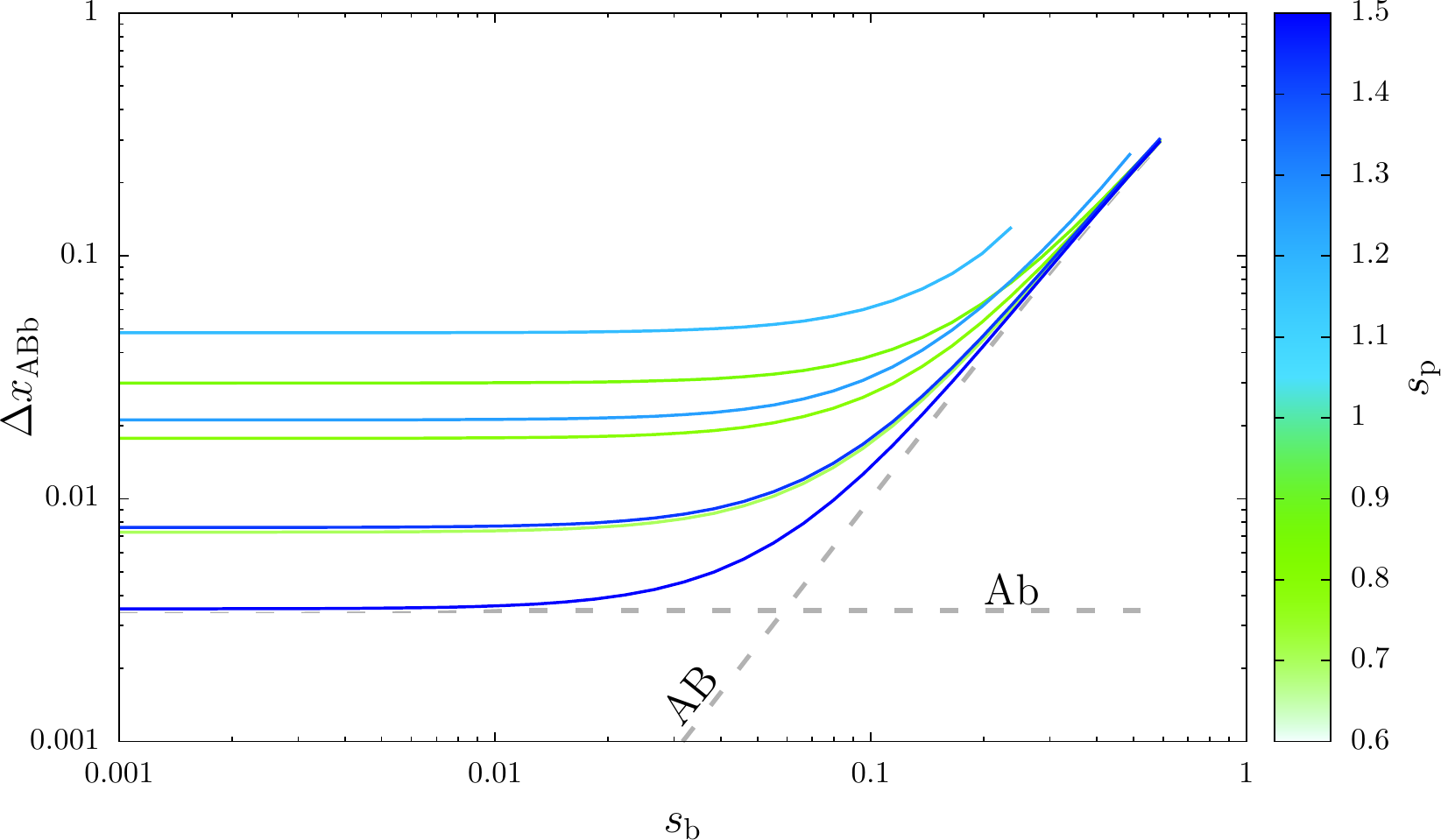}
\includegraphics[width=\columnwidth]{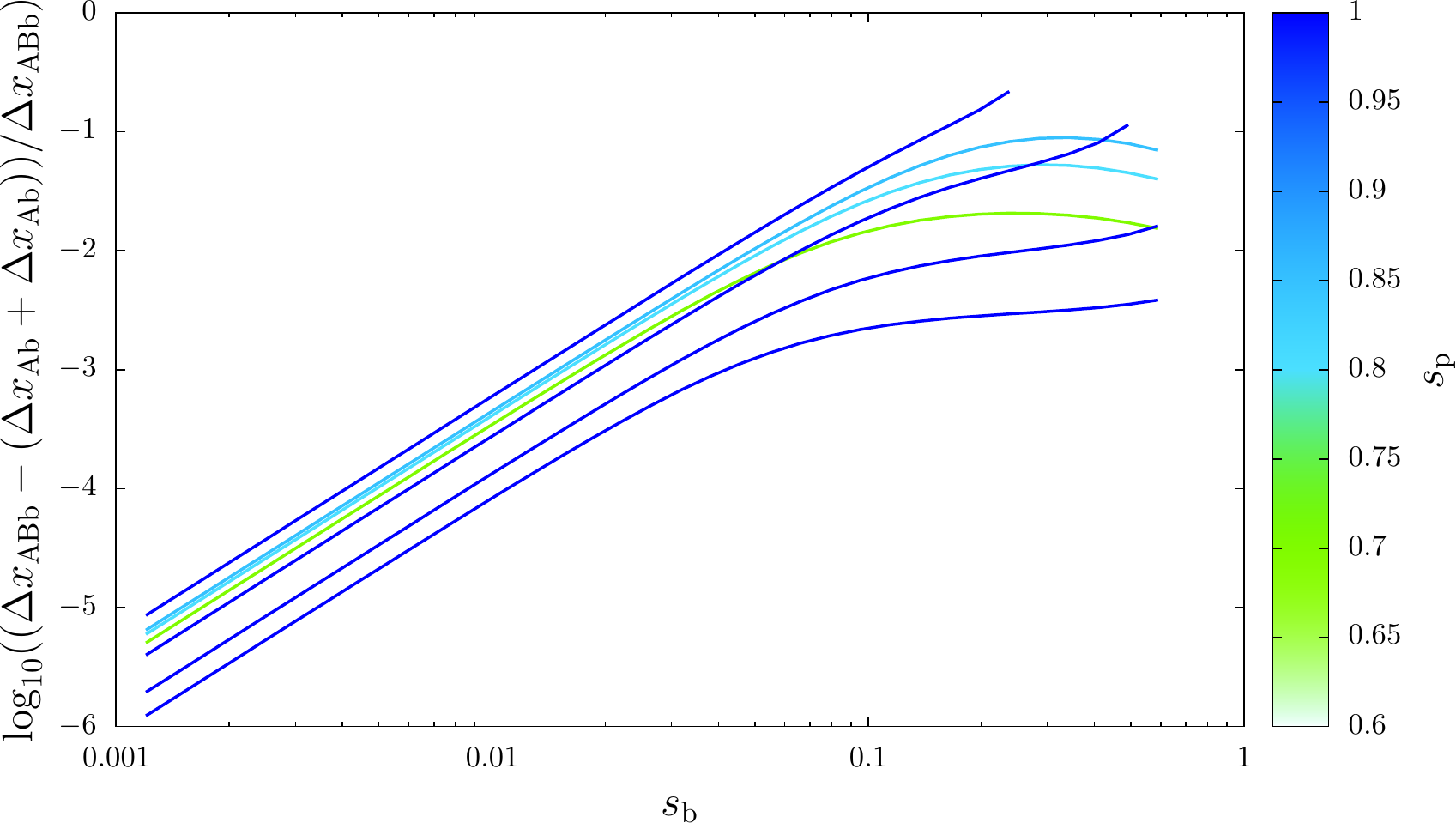}
\caption{Length of the central caustics (for $\phi_{\rm b}$ = 0) as $s_{\mathrm{b}}$ changes for various $s_{\rm p}$, which are encoded by color. The top panel shows the x-axis length of the ABb central caustics for each $s_{\mathrm{b}}$. The two dashed grey lines show the Ab and AB central caustic lengths for $s_{\mathrm{p}}= 0.6$ across the same range of $s_{\mathrm{b}}$. The bottom panel plots the fractional difference between the x-axis length of the ABb central caustic and the sum of the x-axis lengths of the Ab and AB central caustics over the same range of $s_{\mathrm{b}}$. Note that this approximation is good to within $\sim 1\%$ for $s_{\mathrm{b}} < 0.1$ and within $10 \%$ for $s_{\mathrm{b}} < 0.7$. }
\label{central_caustic_length}
\end{figure}

\autoref{central_caustic_length} plots the length (as measured along the $x$-axis) $\Delta x_{\rm ABb}$ of the central caustics for various $s_{\mathrm{p}}$ and $s_{\mathrm{b}}$ values at $\phi_{\rm b} =0$. The top panel shows that, as expected, the length of the central caustic will tend to the length of either the AB or Ab caustic as $s_{\rm p}$ or $s_{\rm b}$ tend to zero, respectively. When the AB and Ab caustics are of a similar size however, the length is larger than either AB or Ab caustic. We approximate the ABb central caustic length as simply the sum of the AB and Ab central caustic lengths. The bottom panel of \autoref{central_caustic_length} plots the fractional error of this approximation and shows that (at least for $\phi_{\rm b}=0$) the approximation that, the lengths of the AB and Ab caustics simply add together to determine the length of the ABb central caustic, holds well. 

\subsection{Planetary Caustics}\label{planetarycaustics}

\subsubsection{Effects of Changing Stellar Binary Angle ($\phi_{\mathrm{b}}$)}\label{phib}

 As $\phi_{\mathrm{b}}$ goes through a full rotation, the center point between the two planetary caustics trace out a shape similar to a lima\c{c}on around the location of the planetary caustic in the Ab system. In the case where $q_{\mathrm{b}} = 1$, the lima\c{c}on approaches the shape of a circle, that is traced out twice for each full rotation of the binary. The planetary caustics of the close topology each follow this general motion, but their lima\c{c}on path is slightly distorted relative to that traced by their center point.

\begin{figure}
\centering
\includegraphics[width=\columnwidth]{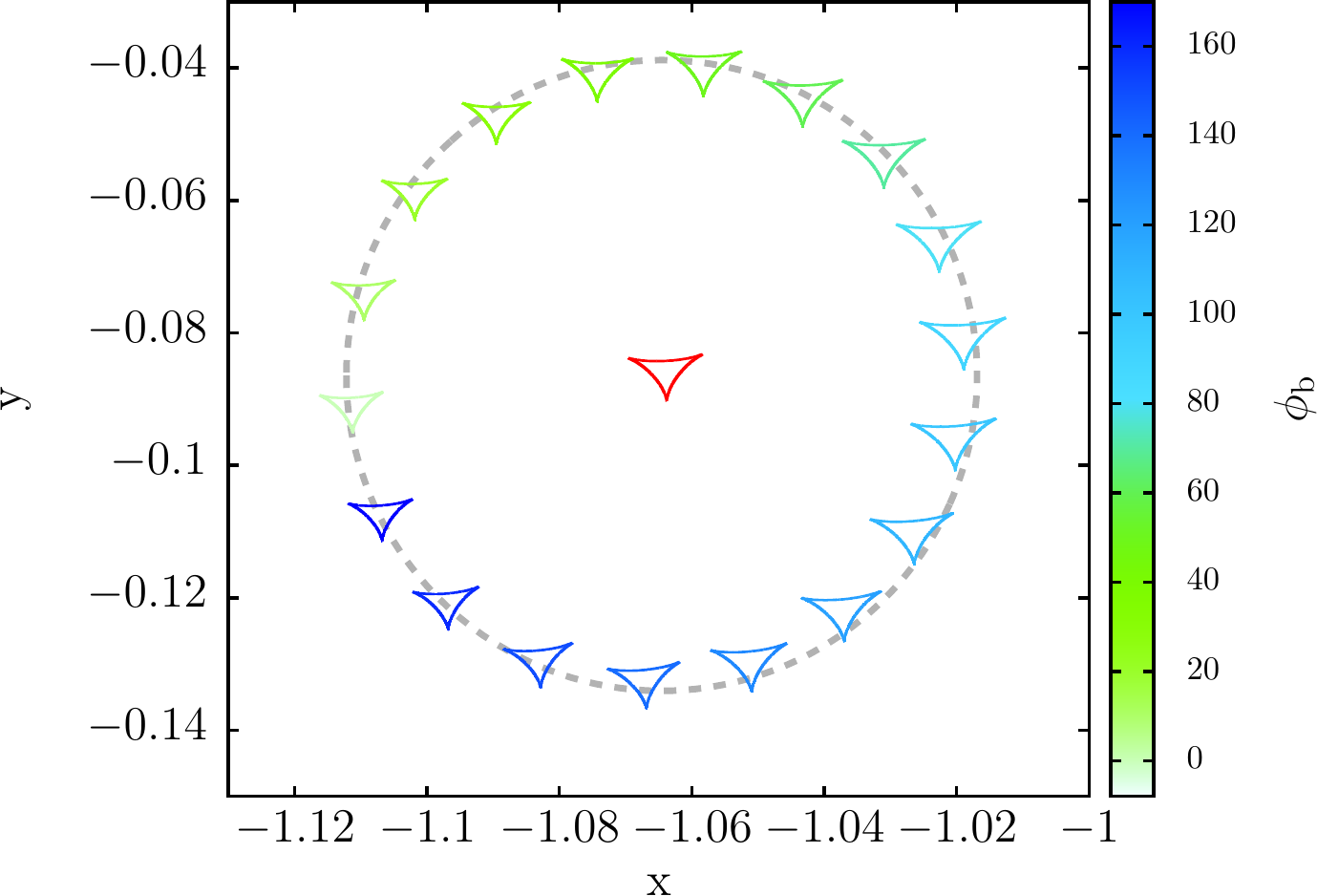} 
\includegraphics[width=\columnwidth]{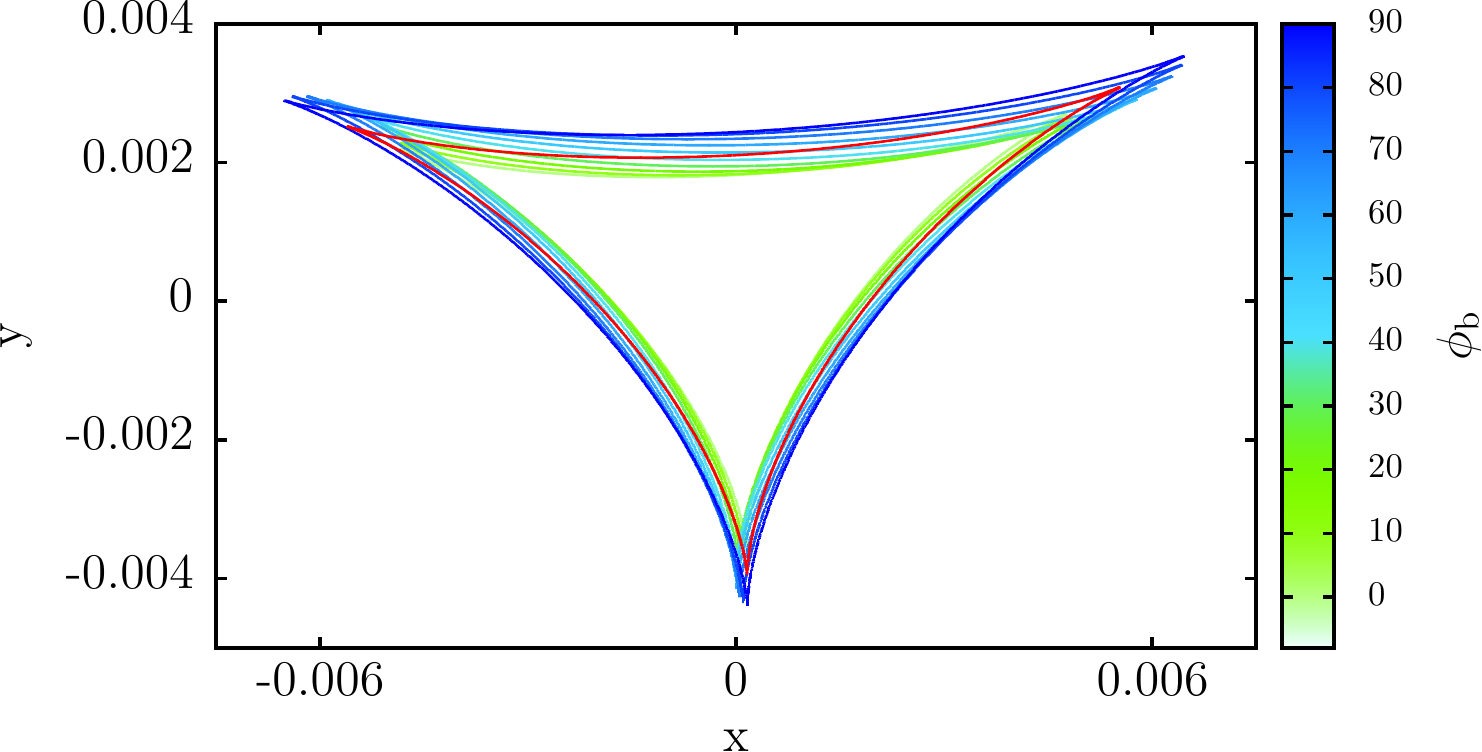} 
\caption{Effect of changing $\phi_{\mathrm{b}}$ on planetary caustics in close topology. The green through blue colored caustics are the triple-lens circumbinary caustics at a progression of values of $\phi_{\mathrm{b}}$. The red caustics show the Ab planetary caustic. The top panel shows the actual caustic positions in the source plane. The bottom panel shows the same caustics overlaid on one another to show the slight change in size and shape. Note that the caustics are displaced by $180^{\circ}$ for stellar binary displacement of $90^{\circ}$. The gray dashed line shows a circle of radius $\delta$ (see \autoref{planetarycausticslocation}).}
\label{revolution1}
\end{figure}

\begin{figure}
\centering
\includegraphics[width=\columnwidth]{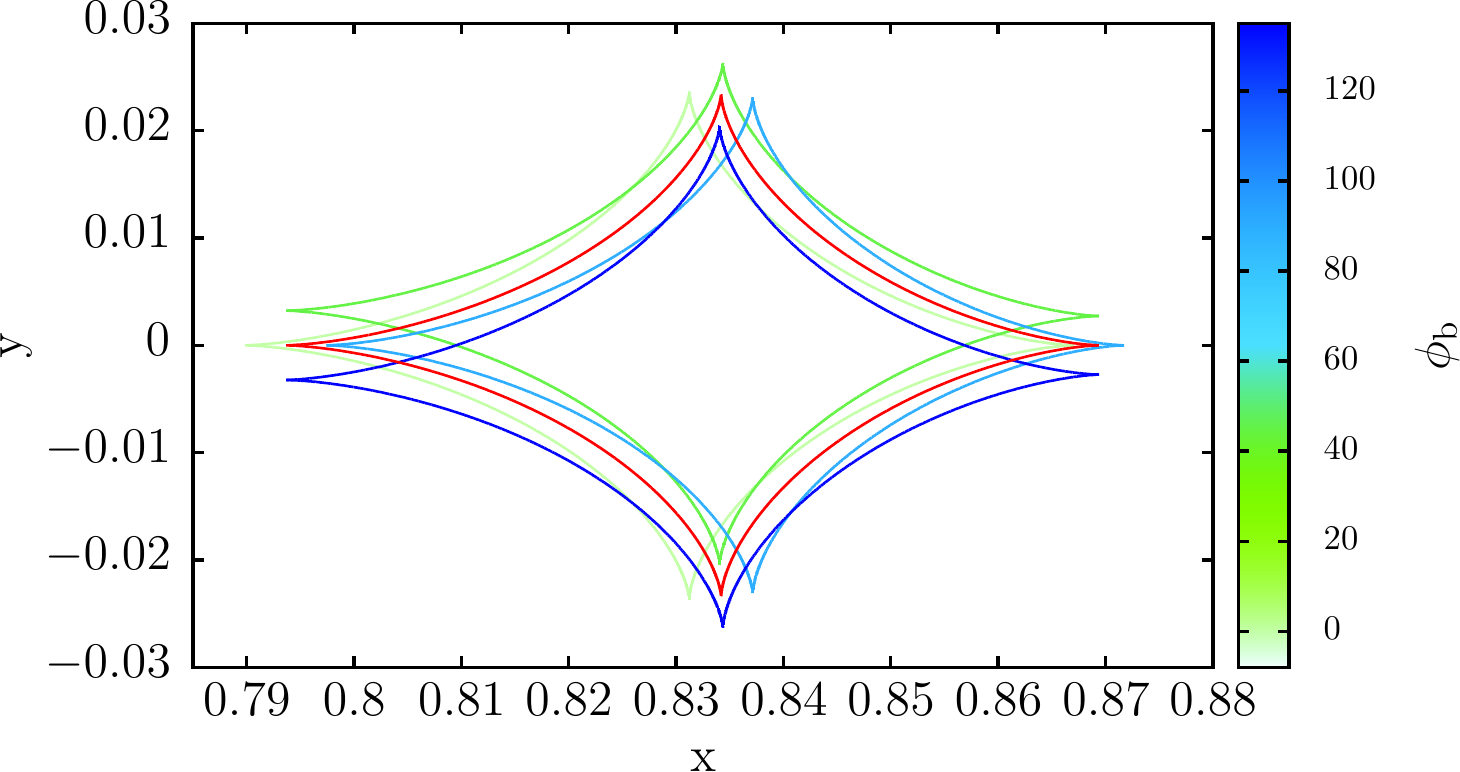} 
\includegraphics[width=\columnwidth]{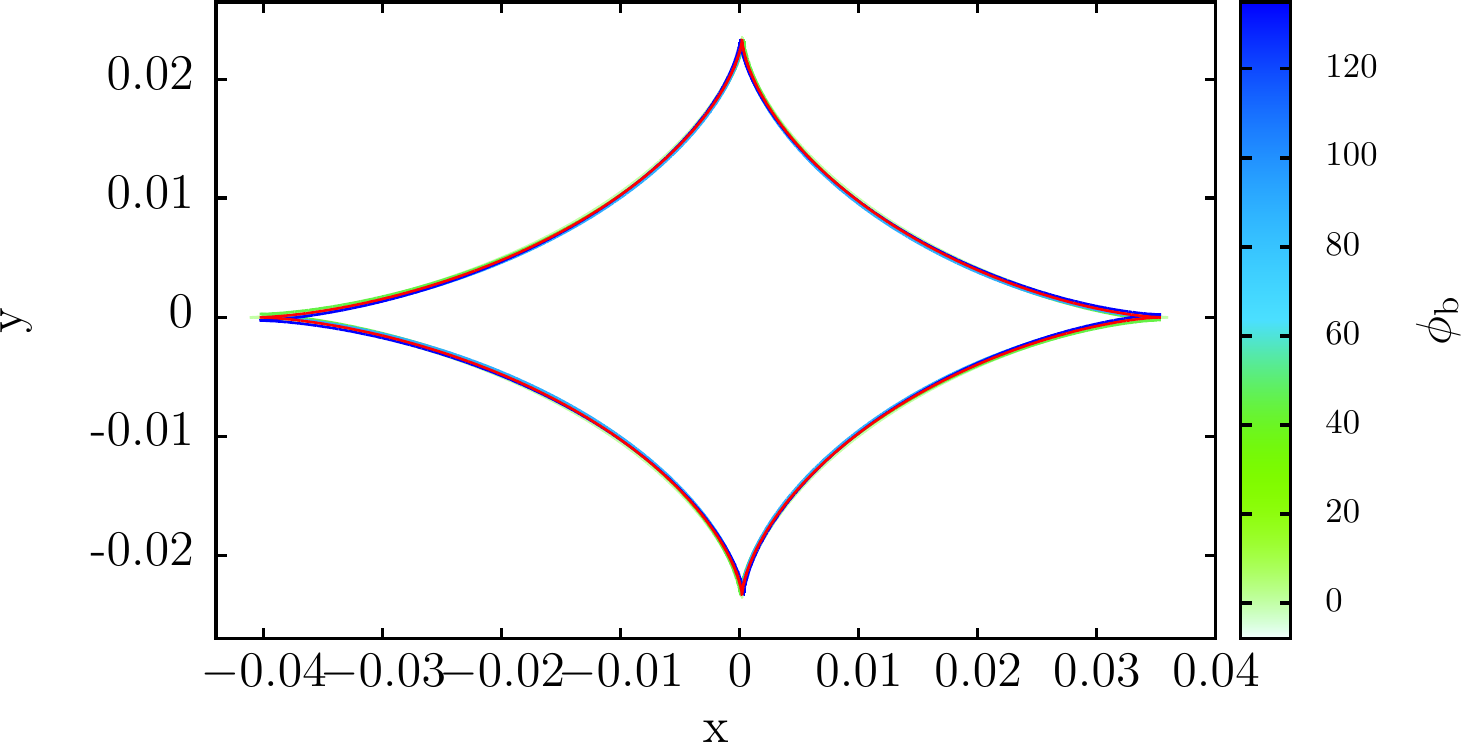} 
\caption{As \autoref{revolution1} but for the wide planetary caustic and changing $\phi_{\rm b}$.}
\label{revolution2}
\end{figure}

The motion of the planetary caustics is evident for the close topology as shown in \autoref{revolution1} and for the wide topology in \autoref{revolution2}. What is not as evident in the figures but is clearly seen in the animations is the anticorrelation between the motion of the caustic and the revolution of the stellar binary: as the stellar binary roates counter-clockwise, the planetary caustic moves clockwise. This motion of the planetary caustics is consistent across the close, and wide topologies. Even in the resonant cases, where the planetary caustics join together with the central caustics to form one caustic, the segments of the caustic that came from the planetary caustics continue to move in this fashion as $\phi_{\mathrm{b}}$ changes. The radius of translation of the planetary caustics is greatest when the planet is in the close topology, and decreases as the planet moves further out. In addition to the translation, there is a slight change in size of the caustics, as shown in the bottom panels of each figure. The relative change in size is largest for close lenses.

\subsubsection{Effects of Changing Stellar Binary Separation ($s_{\mathrm{b}}$)}

\begin{figure}
\centering
\includegraphics[width=\columnwidth]{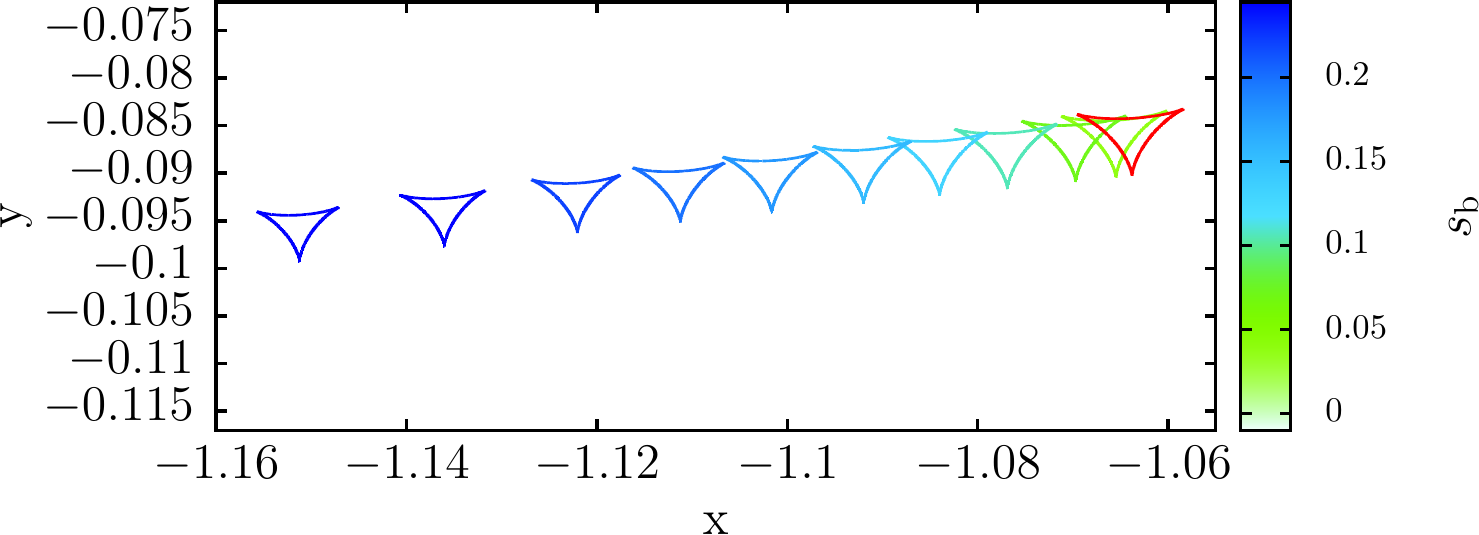} 
\includegraphics[width=\columnwidth]{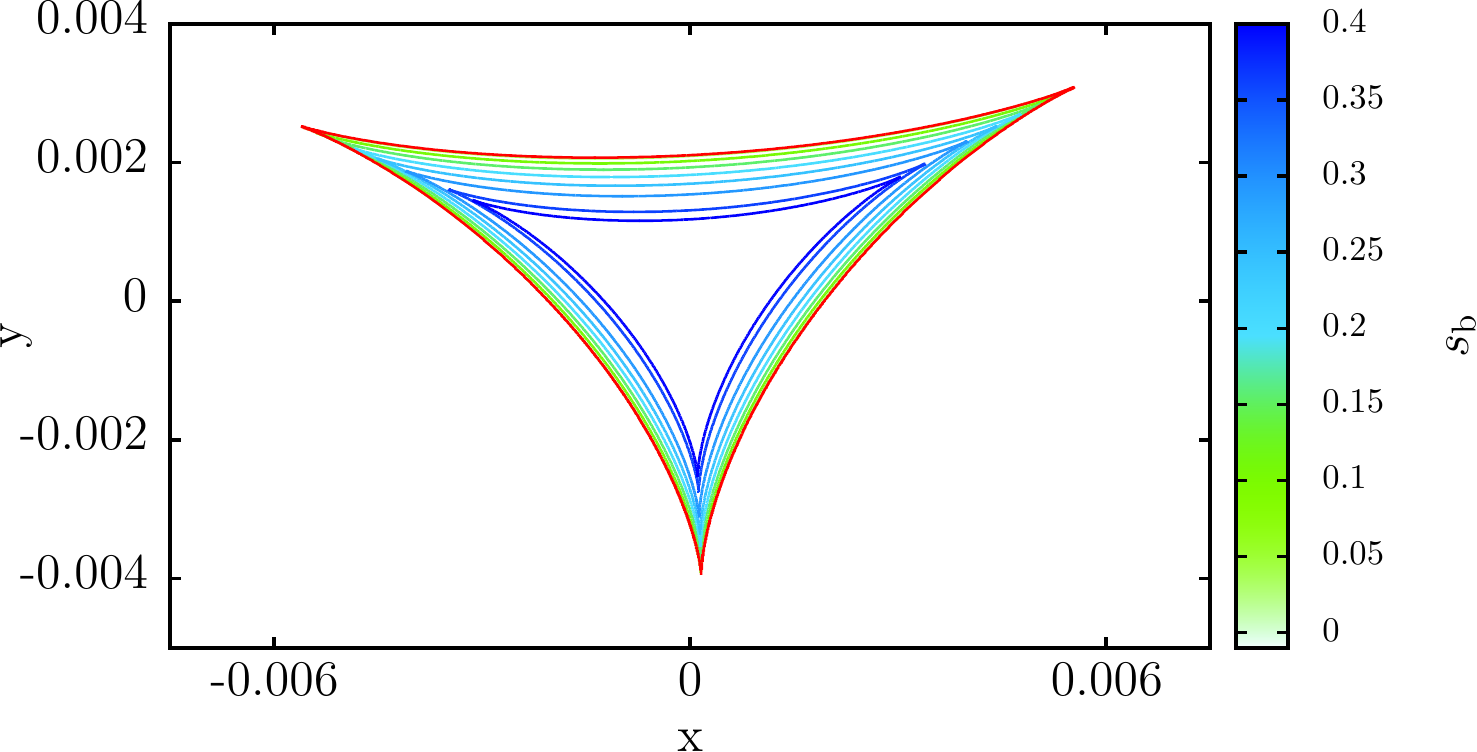} 
\caption{As \autoref{revolution1} but for the close planetary caustic with changing $s_{\rm b}$.  Note that in the top panel, $s_{\mathrm{b}}$ ranges from 0 to 0.2 to show the displacement of the caustics. The bottom panel overlays the caustics for $s_{\mathrm{b}}$ values ranging from $0$ to $0.4$. Note therefore that the color scales differ between the two panels. The difference in size is much greater for changes in $s_{\mathrm{b}}$ than for changes in $\phi_{\mathrm{b}}$.}
\label{translation1}
\end{figure}

\begin{figure}
\centering
\includegraphics[width=\columnwidth]{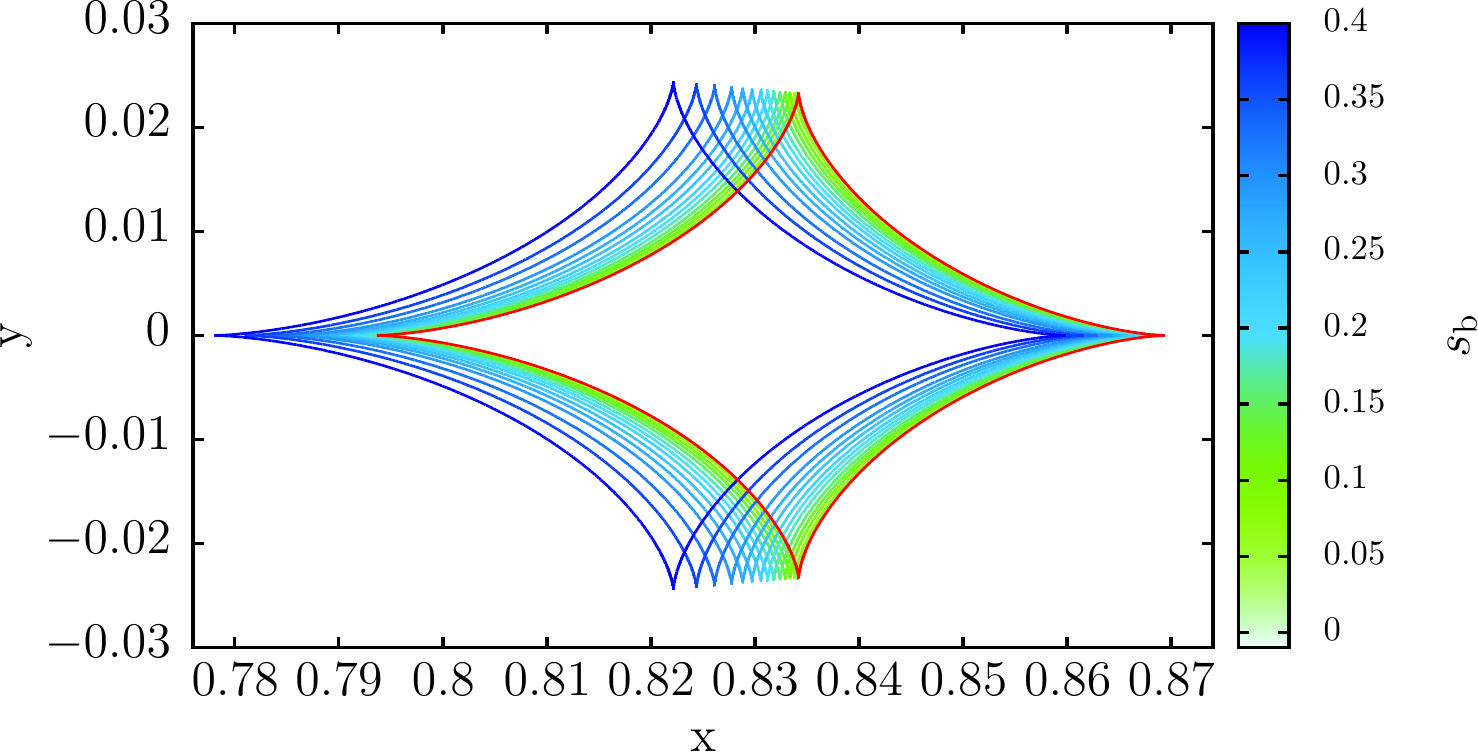} 
\includegraphics[width=\columnwidth]{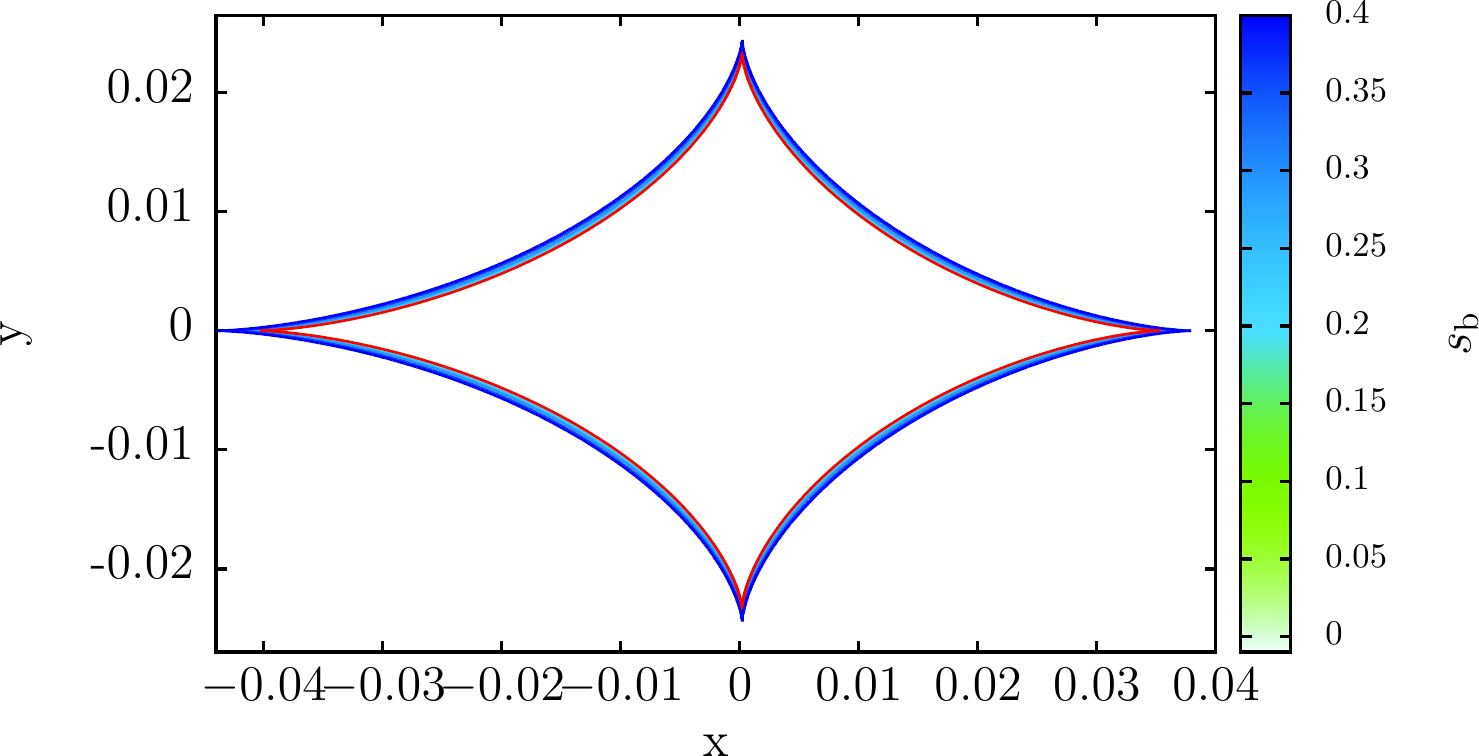} 
\caption{As \autoref{revolution1} but for wide planetary caustics with $s_{\rm b}$ changing. Here, $s_{\mathrm{b}}$ ranges from 0 to 0.4 in both panels. For both changes in $\phi_{\rm b}$ and $s_{\rm b}$ the relative change in size of the wide planetary caustic is much less than the relative change in size of the close planetary caustic.}
\label{translation2}
\end{figure}

Figures~\ref{translation1} and \ref{translation2} show the effect of varying $s_{\rm b}$ on the planetary caustics. When varying $s_{\mathrm{b}}$, the planetary caustics move not in a circle but instead are translated in a roughly straight line. Once again, this movement is based around the planetary caustic position of the Ab system, and when $s_{\rm b}\rightarrow 0$ the planetary caustic position tends to that of the Ab system. In resonant topologies, the position of the caustic itself does not change significantly, but the cusps that were formed from the planetary caustics move and stretch the caustic. In addition to the translational movement, the caustics also change size. As the caustics move farther away from their position at $s_{\mathrm{b}} = 0$, they also shrink in size. The relative change in size is much larger for close topologies.

\subsubsection{Location of the Planetary Caustics}\label{planetarycausticslocation}

\begin{figure}
\includegraphics[width=\columnwidth]{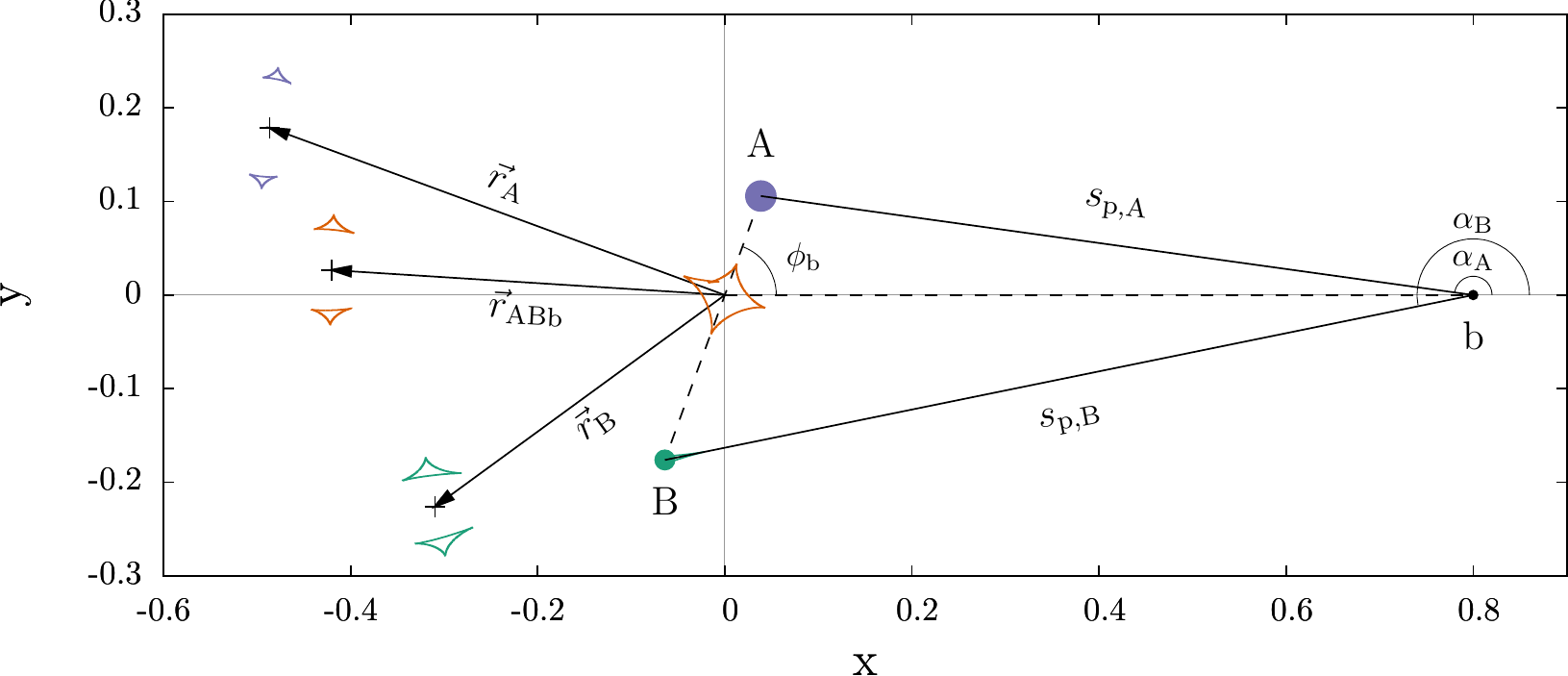} 
\caption[Diagram]
{\singlespace Diagram of ABb caustics, A caustics, and B caustics. The magenta curves represent the caustics produced by star A and the planet. The blue curves represent the caustics produced by star B and the planet. The red curves represent the caustics produced by star A, star B, and the planet. The black crosses mark the central point between the two planetary caustics. For this diagram, $s_{\mathrm{p}} = 0.8$, $s_{\mathrm{b}} = 0.3$, $q_{\mathrm{p}} = 10^{-3}$, $q_{\mathrm{b}} = 0.6$, and $\phi_{\mathrm{b}} = 70^{\circ}$.}
\label{caustics_diagram}
\end{figure}

Inspection of the animations reveals that while the planetary caustics of the ABb system lie close to those of the Ab system, there is some dependence of their position on the parameters of the binary star. A hunch led us to compare the position of the planetary caustics in the ABb system to the positions of the caustics that would be formed by lenses with the planet position fixed but the total mass of the binary star placed either at the position of star A or star B. This scenario can be seen in \autoref{caustics_diagram}. We denote these new systems A and B respectively. To avoid confusion we emphasize the distinction between our systems of interest. \emph{System} A consists of \emph{star} A and the planet b (\autoref{caustics_diagram}). This has the risk of confusion with our earlier notation and references to `Ab' caustics. We reserve `Ab' notation to be the planetary double-lens system where the stellar binary has been replaced by a single star of the same total mass located at the center of mass. Thus, Systems A and B refer to the planetary double-lens system from each star in the binary paired with the planet, whereas System Ab uses a single star as a proxy for both stars in the stellar binary. We found empirically that the location of the ABb planetary caustics $\rabb$ could be accurately predicted by taking a mass weighted average of the (numerically computed) positions of the planetary caustics in the A and B systems, $\bm{r}_{\rm A}$ and $\bm{r}_{\rm B}$ respectively, i.e.,
\begin{equation}
\rabb \simeq \epsilon_{\rm A}\bm{r}_{\rm A} + \epsilon_{\rm B}\bm{r}_{\rm B},
\label{tripleposition}
\end{equation}
where $\epsilon_{\rm A} = m_{\rm A}/(m_{\rm A}+m_{\rm B}) =1/(1+q_{\rm b})$ and $\epsilon_{\rm B} = m_{\rm B}/(m_{\rm A}+m_{\rm B})=q_{\rm b}/(1+q_{\rm b})$ are the mass of stars A and B normalized to the total mass of the binary star and $q_{\mathrm{b}}$ is the ratio of the two stars in the binary. Note that we use $q$'s to describe mass ratios of two objects and $\epsilon$'s to describe mass fractions of the total mass.

We can derive an approximate analytic expression for $\rabb$ by utilizing the analytic approximations of \citet{Bozza2000-special} and \citet{Han2006} and some simple geometry. For a given projected separation between a star and planet $s$, and a mass ratio $q$, the position of the planetary caustic(s) relative to the star is \citep[using the slightly more accurate expression of][]{Bozza2000-special}
\begin{equation}
\bm{r} \simeq \left[\left(\frac{1-q}{1+q}\right)\left(s-\frac{1}{s}\right),0\right],
\label{genericcausticpos}
\end{equation}
where for the close topology, this vector is the position of the center point between the two planetary caustics.
We define the projected separations of the A and B systems as $s_{\mathrm{p,A}}$ and $s_{\mathrm{p,B}}$, respectively. Utilizing the fact that the center of the A/B planetary caustics will lie along the vector between star A/B and the planet, the $x$ and $y$ components of the $\bm{r}_{\rm A}$ vector can be written
\begin{subequations}
\begin{align}
x_{\mathrm{A}} & \simeq s_{\mathrm{p}} + \left\{s_{\mathrm{p,A}} - \left(\frac{1-q_{\mathrm{p}}}{1+q_{\mathrm{p}}}\right)\left(s_{\mathrm{p,A}} - \frac{1}{s_{\mathrm{p,A}}}\right)\right\}\cos(\alpha_{\mathrm{A}}), \label{causticpositionsx}\\
y_{\mathrm{A}} & \simeq \left\{s_{\mathrm{p,A}} - \left(\frac{1-q_{\mathrm{p}}}{1+q_{\mathrm{p}}}\right)\left(s_{\mathrm{p,A}} - \frac{1}{s_{\mathrm{p,A}}}\right)\right\}\sin(\alpha_{\mathrm{A}}),
\label{causticpositionsy}
\end{align}
\end{subequations}
and similarly for the components of $\bm{r}_{\rm B}$, with each A replaced with a B. The angles $\alpha_{\rm A}$ and $\alpha_{\rm B}$ are the angles subtended by the binary star components as viewed from the planet and measured from the line connecting the planet to the binary center of mass (see \autoref{caustics_diagram}). They can be written as
\begin{equation}
\alpha_{\rm A} = \pi - {\rm atan2}\left(\epsilon_{\rm B}s_{b}\sin{\phi_{\rm b}},s_{\rm p} - \epsilon_{\rm B}s_{b}\cos{\phi_{\rm b}}\right),
\end{equation}
and
\begin{equation}
\alpha_{\rm B} = \pi + {\rm atan2}\left(-\epsilon_{\rm A}s_{b}\sin{\phi_{\rm b}},s_{\rm p} + \epsilon_{\rm A}s_{b}\cos{\phi_{\rm b}}\right),
\end{equation}
using the convention of ${\rm atan2}(y,x)$ for the order of arguments.
By the cosine rule, the projected separations of the planet relative to each star A and B are
\begin{equation}
s_{\mathrm{p,A}} = \sqrt{s_{\mathrm{p}}^2 - 2 \epsilon_{\rm B} s_{\mathrm{p}} s_{\mathrm{b}} \cos(\phi_{\mathrm{b}}) + \epsilon_{\rm B}^2 s_{\mathrm{b}}^2} \,,
\label{spa}
\end{equation}
\begin{equation}
s_{\mathrm{p,B}} = \sqrt{s_{\mathrm{p}}^2 + 2 \epsilon_{\rm A} s_{\mathrm{p}} s_{\mathrm{b}} \cos(\phi_{\mathrm{b}}) + \epsilon_{\rm A}^2 s_{\mathrm{b}}^2} \,,
\label{spb}
\end{equation}
respectively.

\begin{figure}
\includegraphics[width=\columnwidth]{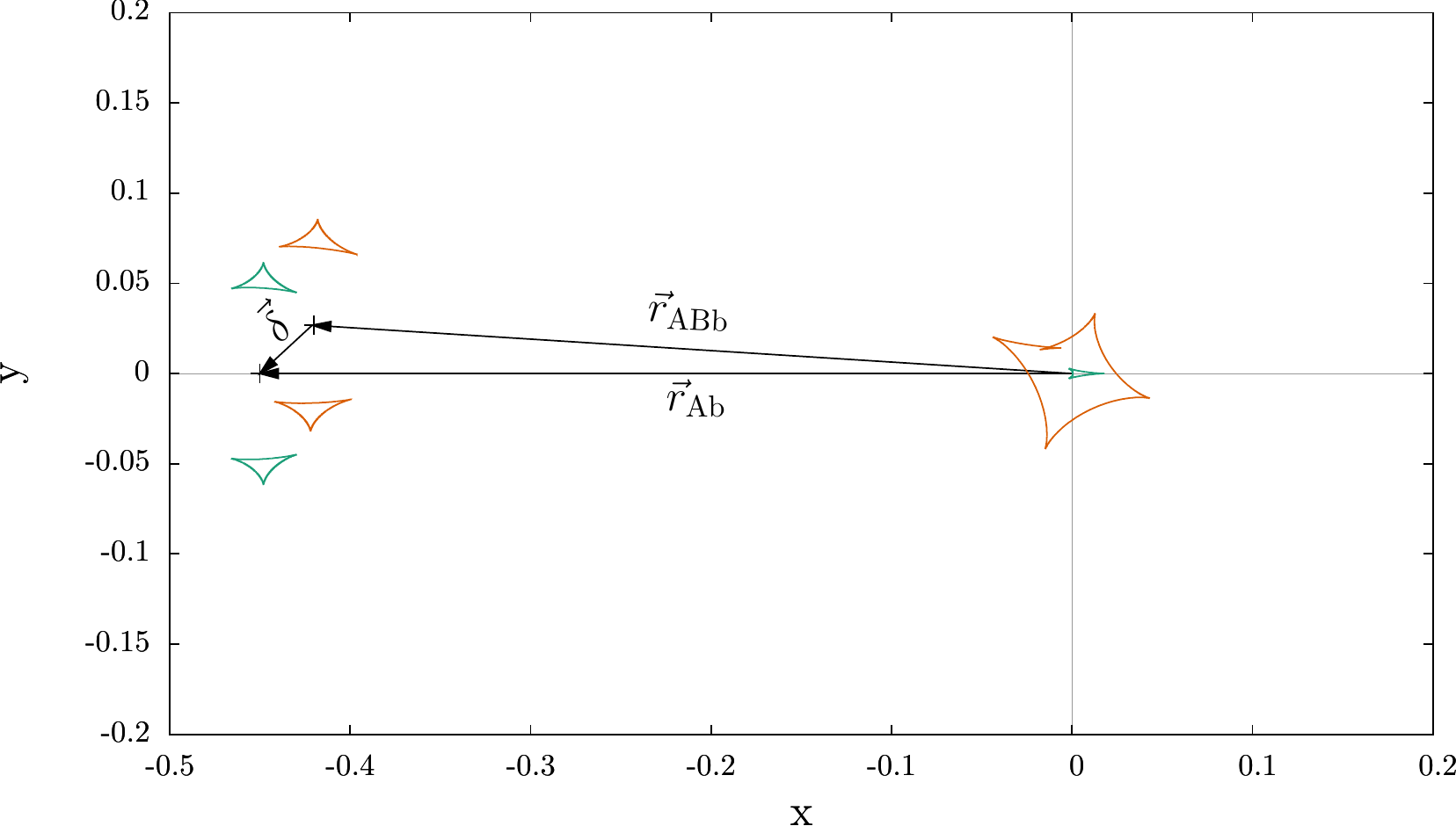}
\caption{Diagram showing the ABb planetary caustics (orange), Ab planetary caustics (green), and the vector $\mathbf{\delta}$ between them, as defined in \autoref{dvector}. Parameters as in \autoref{caustics_diagram}.}
\label{delta_plot}
\end{figure}

In general, the expression for $\rabb$ is analytically cumbersome, but can be quickly computed numerically. However, we can make significant progress analytically by considering a special case and using some observations from the animations. It is useful at this point to define the vector
\begin{equation}
\bm{\delta} \equiv \rabb - \rab,
\label{dvector}
\end{equation}
which is the displacement of the center point of the planetary caustic(s) in the ABb system from that in the Ab system. \autoref{delta_plot} depicts a diagram showing this vector.
From the animations, we can see that the displacement of each planetary caustic for close lenses is always in the same sense, as if the caustics were connected by a bar that must remain vertical. This means that $\bm{\delta}$ approximately describes the displacement of the individual planetary caustics for close lenses, as well as the center point between them. As can be seen in the animations, as $\phi_{\rm b}$ changes, the caustics move in a lima\c{c}on, which is reasonably approximated by a circle. The caustics complete two full rotations for every rotation of the binary, so we can write $\bm{\delta}$ as a vector in polar coordinates with the origin at the location of the Ab caustic
\begin{equation}
\bm{\delta} \simeq (\delta,-2\phi_{b}),
\label{dpolar}
\end{equation}
where the negative sign again indicates the anticorrelation between stellar binary rotation and the movement of the caustics. We can therefore compute $\delta$ and have a reasonable approximation for the position of the ABb planetary caustics for any $\phi_{\rm b}$. For the case of a planet ($q_{\rm p}\ll1$) orbiting an equal mass binary star ($q_{\rm b}=1$), when $\phi_{\rm b}=0$, $y_{\rm ABb}=0$ and the expression for $x_{\rm ABb}$ depends only on $s_{\rm b}$ and $\phi_{\rm b}$. With a little algebra, the expression for $\delta$ can be written as
\begin{equation}
\delta \simeq \frac{s_{\rm b}^2}{4 s_{\rm p}^3}\left\lbrack 1 - \frac{1}{4}\left(\frac{s_{\rm b}}{s_{\rm p}}\right)^2\right\rbrack^{-1}.
\label{dequation}
\end{equation}
While we will not attempt to derive a simplified version of the more general analytic expression for $\bm{\delta}$, it is worth noting that accurate estimates for the positions of planetary caustics in general can be computed using Equations~\ref{tripleposition} and \ref{causticpositionsx} through \ref{spb}, without the need to numerically solve the sixth order polynomial caustic equations.

\section{Orbital Motion of the Stellar Binary}\label{orbitalmotion}

The previous section has highlighted a number of features that can be observed in the animations of the ABb caustics. While the animations are mainly meant to give a more intuitive understanding of how the ABb caustis depend on the stellar binary parameters $\phi_{\mathrm{b}}$ and $s_{\mathrm{b}}$, they can also be used as crude simulations of the orbital motion of the stellar binary. With this in mind, it is clear that the orbital motion of the stellar binary can induce significant changes in the shape and/or position of the planetary caustic. In this section we consider the effect of the binary's orbital motion on the caustics and compare it to the effect of the planet's orbital motion.

In some planetary microlensing events, the time it takes the source to cross the Einstein radius can be long enough that the orbital motion of the planet can be detected due to the change it induces on the caustic features~\citep[e.g.][]{Dominik1998, Ioka1999, Albrow2000,Penny2011-om}. With this fact in mind, after looking at how the stellar binary of our triple-lens system can affect the shape, size, and position of the caustics, it is reasonable to ask if the orbital motion of the stellar binary could be detected due to the motion it induces on the planetary caustics, or if the caustic motion may be confused for planetary orbital motion. After all, the stellar binary is on a closer orbit than the planet, and would therefore have a faster angular velocity.

We quantify the question by computing the ratio of the speed of the planetary caustic induced by orbital motion of the binary star $|\rabbdot|$ to the speed of the planetary caustic from the planet's orbital motion $|\rabdot|$, which we define as
\begin{equation}
\Omega \equiv \frac{|\rabbdot|}{|\rabdot|}
\label{ratioeqn}
\end{equation}
The time derivatives in this equation will depend on the orientation of the orbit, which in general will be inclined and/or eccentric, but for simplicity, we examine only the cases of face- and edge-on circular orbits that are not mutually inclined.

\subsection{Face-on Orbits}
When face-on, the orbital motion of the stellar binary corresponds to $\phi_{\mathrm{b}}$ changing with $s_{\rm b}$ fixed, and the orbital motion of the circumbinary planet corresponds to a rotation of the whole frame, which we assume we can approximate with the motion of the planetary caustic in the Ab system. 

With the chain rule, we can write the time derivatives in Equation \ref{ratioeqn} as
\begin{equation}
|\rabbdot| = \left|\frac{\dd\rabb}{\dd\phi_{\rm b}}\right| \omega_{\rm b},
\end{equation}
and
\begin{equation}
|\rabdot| = \left|\frac{\dd\rab}{\dd\phi_{\rm p}}\right| \omega_{\rm p},
\end{equation}
where $\omega_{\rm b}$ and $\omega_{\rm p}$ are the angular speeds of the binary and planet, respectively.
In the Ab system, the caustic simply rotates around the center of mass, so $|\dd\rab/\dd\phi_{\rm p}|$ is a constant
\begin{equation}
\left|\frac{\dd\rab}{\dd\phi_{\rm p}}\right| = |\rab|\ \text{rad}^{-1} \simeq \left|s_{\rm p}-\frac{1}{s_{\rm p}}\right|\ \text{rad}^{-1}.
\end{equation}
If we assume the mass of the planet is negligible compared to the mass of the binary, the ratio of angular speeds depends only on the ratio of separations and can be written
\begin{equation}
\frac{\omega_{\mathrm{b}}}{\omega_{\mathrm{p}}} = \left(\frac{s_{\mathrm{b}}}{s_{\mathrm{p}}}\right)^{-\frac{3}{2}},
\label{omegaeqn}
\end{equation}
using Kepler's third law.
From the animations in \autoref{animations}, we can see that for all caustic topologies the ABb planetary caustics translate approximately in a circle around the Ab planetary caustics, completing two rotations for every complete rotation of the binary. We can therefore write 
\begin{equation}
\left|\frac{\dd\rabb}{\dd\phi_{\mathrm{b}}}\right| \simeq 2|\rabb-\rab|\ \text{rad}^{-1} \approx 2\delta\ \text{rad}^{-1},
\end{equation}
where $\delta$ is the radius of the circle transcribed by the ABb caustic, an expression for which we derived in \autoref{dequation}.

\begin{figure}
\includegraphics[width=\columnwidth]{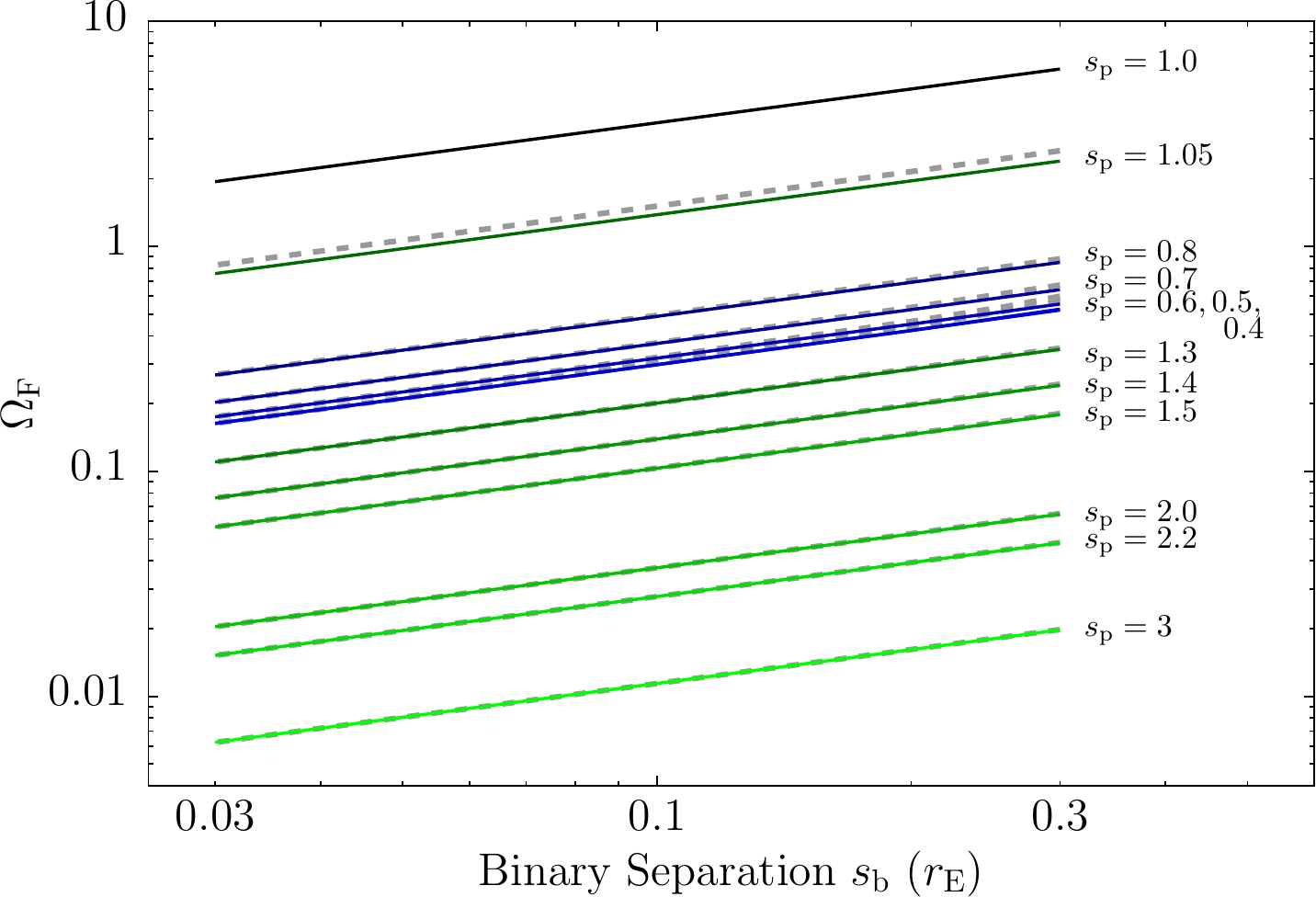} \\ 
\includegraphics[width=\columnwidth]{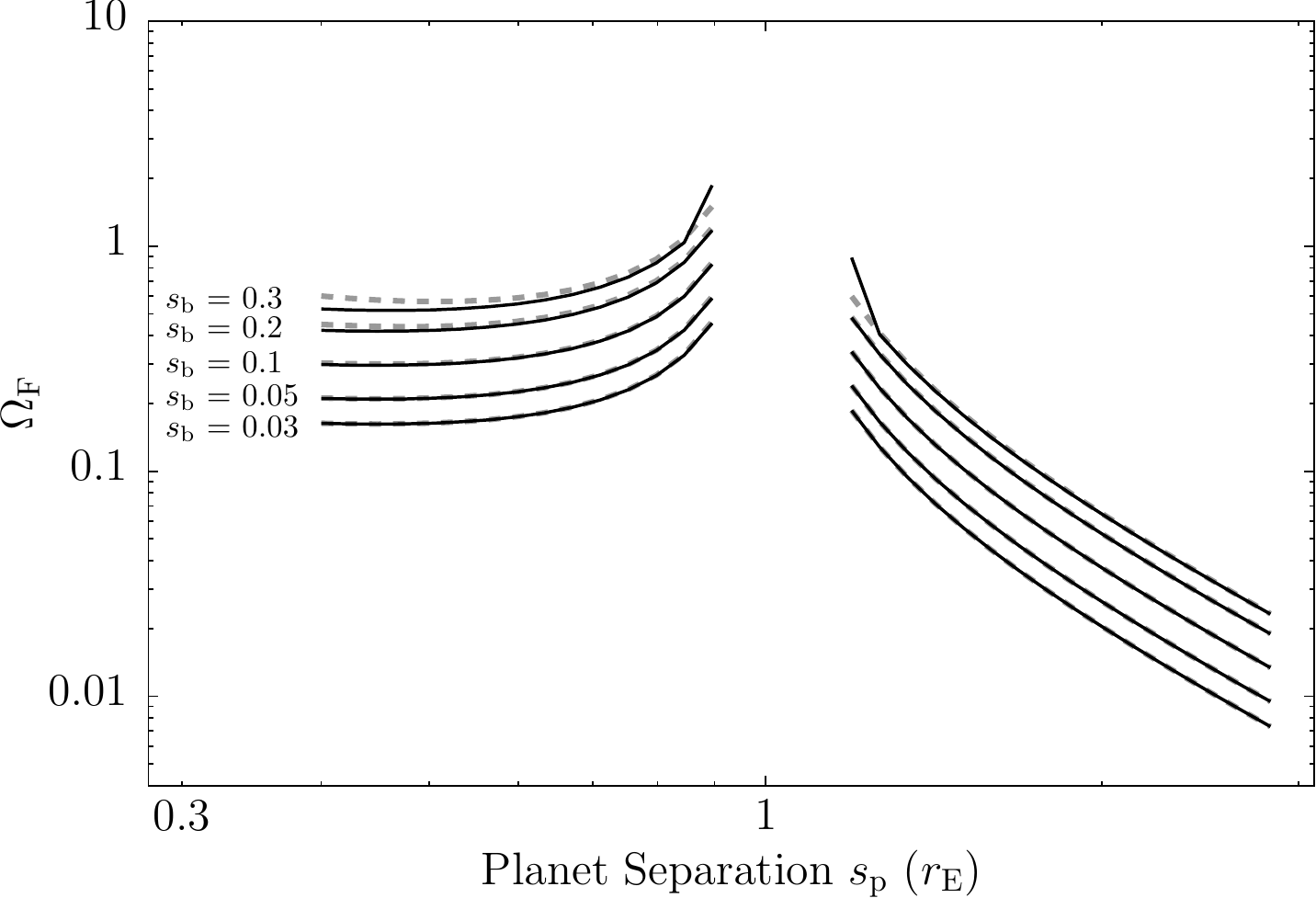}
\caption{Ratio of the ABb planetary caustic speed caused by the binary's orbital motion to the Ab planetary caustic speed caused by the planet's orbital motion for a face-on orbit. The top panel plots $\Omega_{\mathrm{F}}$ as a function of $s_{\mathrm{b}}$ for various $s_{\mathrm{p}}$ values, and the bottom panel plots the same data as a function of $s_{\mathrm{p}}$ for the various $s_{\mathrm{b}}$ values. Solid lines indicate the numerically computed values. In the top panel, the colors help to distinguish the $s_{\mathrm{p}}$ values. Green indicates planetary separations greater than 1 (wide topology) and blue indicates planet separations less than $1$ (close topology), with darker shades being closer to $s_{\rm p}=1$. The hardly-visible thick grey dashed lines underplotted in both panels show the analytical results from Equation \ref{face-on2}, however it is not plotted for $s_{\mathrm{p}} = 1$, since the approximation for the location of planetary caustics ($s_{\mathrm{p}}-1/s_{\mathrm{p}}$) breaks down and $\Omega_{\mathrm{F}}$ is infinite.}
\label{caustic_motion1}
\end{figure}

Combining each of these ingredients, we can write the ratio of caustic speeds for face-on orbits as
\begin{equation}
\Omega_{\rm F} \approx \frac{2 \delta}{|s_{\rm p}-1/s_{\rm p}|} \left(\frac{s_{\mathrm{b}}}{s_{\mathrm{p}}}\right)^{-\frac{3}{2}},
\label{face-on}
\end{equation}
where the subscript F signifies the face-on orbit. Incorporating the expression for $\delta$ when $q_{\rm b}=1$ in \autoref{dequation} and some algebra, this can be rearranged to
\begin{equation}
\Omega_{\rm F} \approx \half \mfracpow{s_{\rm p}}{3}{2} s_{\rm b}\phalf \left|s_{\rm p}-\frac{1}{s_{\rm p}}\right|^{-1} \left\lbrack 1-\frac{1}{4}\left(\frac{s_{\rm b}}{s_{\rm p}}\right)^2\right\rbrack^{-1}.
\label{face-on2}
\end{equation}
For small $s_{\rm b}$ (which is always smaller than $s_{\rm p}$ for face-on orbits), $\Omega_{\rm F}$ scales proportional to $s_{\rm b}^{1/2}$, which implies that motion of the caustics is maximized by increasing the strength of perturbations due to the binary rather than increasing the angular velocity of the binary. For small $s_{\rm p}\ll 1$, $\Omega_{\rm F}\propto s_{\rm p}^{-1/2}$, but this limiting behavior is only valid for $s_{\rm p}$ values that correspond to undetectable planetary caustics, and so we have not shown this behavior in \autoref{caustic_motion1}. For $s_{\rm p}\gg 1$, $\Omega_{\rm F}$ falls off steeply proportional to $s_{\rm p}^{-5/2}$. Inspection of \autoref{face-on2} shows that, because $s_{\rm p}>s_{\rm b}$, all the terms except the $|s_{\rm p}-1/s_{\rm p}|$ term are always smaller than $1$. Therefore, the only way in which the binary's motion can induce caustic motion that is faster than that induced by the planet's orbital motion is when $s_{\rm p}\approx 1$, i.e., when the caustic is resonant. However, it is when the resonant caustic is encountered that the circumbinary nature of the lens is most likely to be recognized. Therefore, in the case of face-on orbits, the motion of the caustic caused by the circumbinary host is unlikely to be confused for planetary orbital motion. 

\autoref{caustic_motion1} shows \autoref{face-on2} as a function of $s_{\rm b}$ and $s_{\rm p}$, together with values of $\Omega_{\rm F}$ computed numerically.  There is very good agreement between the analytic and numerical results -- the analytic lines are plotted in gray beneath the numerical lines. While the planet-induced motion is almost always larger than the binary-induced motion, the binary-induced motion is still significant for larger binary separations, and should be accounted for in any modeling of circumbinary lenses. For the numerical calculation, we estimated $|\dd\rabb/\dd\phi_{\mathrm{b}}|$ by computing the average distance traveled by the center of the ABb planetary caustic per degree change in $\phi_{\mathrm{b}}$ for a large number of $\phi_{\rm b}$ values; the results with a step size of $1$~degree were numerically stable. The center of the caustic was taken to be the average position of caustic points equally spaced in the parameter $\psi$ in \autoref{criticalcurvesoln}, which means the samples will cluster near the cusps of the caustic and the position will be weighted roughly by the magnification structure near the caustic. 

\subsection{Edge-On Orbits}

The edge-on scenario is significantly more complicated than the face-on scenario, as now both $s_{\rm b}$ and $s_{\rm p}$ are a function of time. This means that $\Omega_{\rm E}$ (the subscript E now indicates the edge-on system) will be a function of time, which is more complicated than is desirable for our simple order of magnitude calculation. However, we can perform a dimensional analysis-style calculation and still get an idea of the relative importance of the binary's and planet's motion and how these scale. 

We follow the same steps as in the last subsection and replace any time-varying quantity with its orbit-averaged value. Now, the time derivatives in \autoref{ratioeqn} can be written
\begin{equation}
|\rabbdot| \approx \left|\frac{\dd\rabb}{\dd s_{\rm b}}\right| \langle \dot{s}_{\rm b}\rangle,
\label{rabbdots}
\end{equation}
and
\begin{equation}
|\rabdot| \approx \left|\frac{\dd\rab}{\dd s_{\rm p}}\right| \langle \dot{s}_{\rm p}\rangle.
\label{rabdots}
\end{equation}
Under the same simplifications as the previous subsection ($q_{\rm b}=1$ nd $q_{\rm p}\ll1$) the derivatives with respect to $s_{\rm b}$ and $s_{\rm p}$ are easily calculated to be
\begin{equation}
\left|\frac{\dd \rabb}{\dd s_{\rm b}}\right| \simeq \frac{1}{4}\left\lbrack\left( s_{\rm p}-\frac{s_{\rm b}}{2}\right)^{-2} - \left( s_{\rm p}+\frac{s_{\rm b}}{2}\right)^{-2}\right\rbrack,
\end{equation}
and
\begin{equation}
\left|\frac{\dd \rab}{\dd s_{\rm p}}\right| \simeq 1 + \frac{1}{s_{\rm p}^2}.
\end{equation}
Remembering that the projected separation is always positive, the average of the time derivatives are
\begin{equation}
\langle \dot{s}_{\rm b}\rangle = \frac{2 a_{\rm b} \omega_{\rm b}}{\pi\re},
\end{equation}
where $a_{\rm b}$ is the semimajor axis of the binary; the expression for $\langle\dot{s}_{\rm p}\rangle$ is the same but with subscripts b exchanged for p. The semimajor axis of both the binary and the planet are proportional to the typical projected separation of each, so the ratio will scale as
\begin{equation}
\frac{\langle \dot{s}_{\rm b}\rangle}{\langle \dot{s}_{\rm p}\rangle} \sim \left(\frac{s_{\rm b}}{s_{\rm p}}\right)^{-\frac{1}{2}}.
\label{sdotratio}
\end{equation}
Combining these, and with a little algebra we arrive at the expression
\begin{equation}
\Omega_{\rm E} \sim \frac{1}{2} \powhalf{s_{\rm b}} \mpowhalf{s_{\rm p}}  \left(1+s_{\rm p}^2\right)^{-1} \left\lbrack 1-\frac{1}{4}\left(\frac{s_{\rm b}}{s_{\rm p}}\right)^2\right\rbrack^{-2}.
\label{edge-on}
\end{equation}

\begin{figure}
\includegraphics[width=\columnwidth]{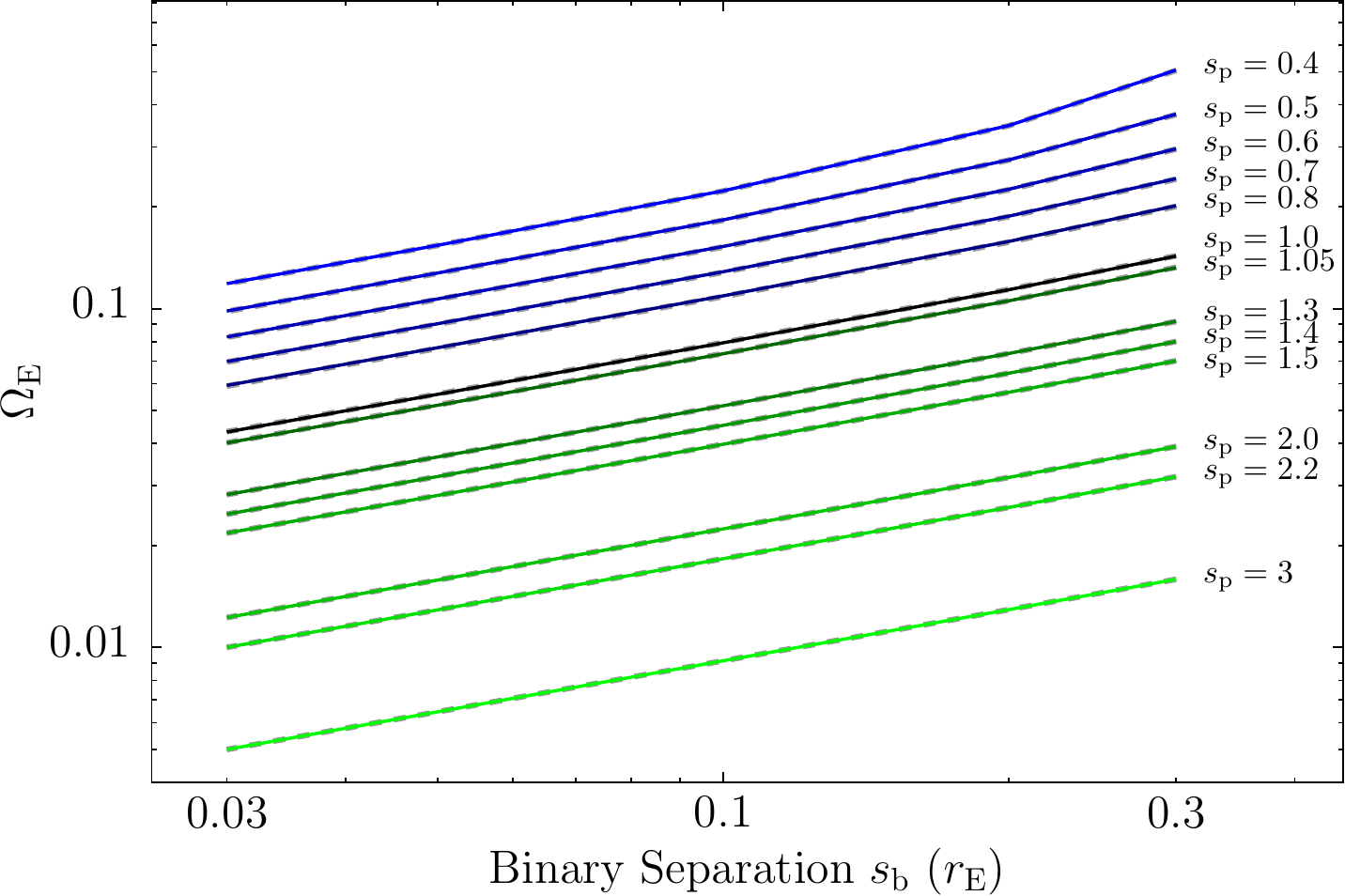} 
\includegraphics[width=\columnwidth]{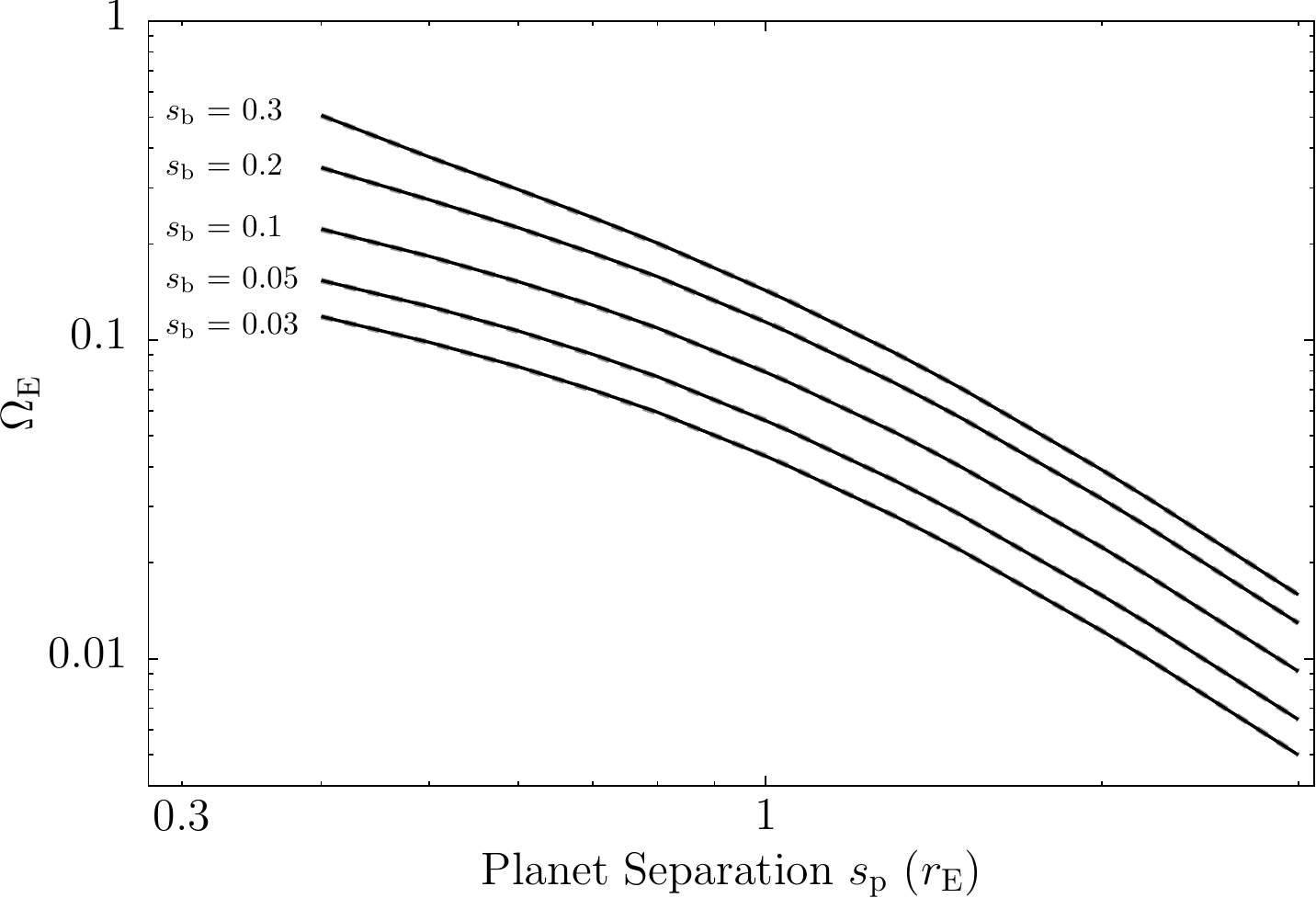}
\caption{As \autoref{caustic_motion1} but for the edge-on caustic speed ratio $\Omega_{\rm E}$.}
\label{caustic_motion2}
\end{figure}

$\Omega_{\rm E}$ has the same limiting behaviour as $\Omega_{\rm F}$. However, wheras $\Omega_{\rm F}$ has a singularity at $s_{\rm p}=1$, $\Omega_{\rm E}$ has a singularity at $s_{\rm p}=s_{\rm b}/2$. For this calculation however, we have not clearly defined the projected separations, so this should not be overinterpreted. The motion of either caustic can vanish when the orbit is at the appropriate phase, and so even if on average the planet's orbital motion induces the majority of caustic motion, at certain times the binary can be the larger contributor. \autoref{caustic_motion2} shows the functional form of $\Omega_{E}$ for a range of $s_{\rm b}$ and $s_{\rm p}$ values, calculated both numerically and analytically using \autoref{edge-on}. The numerical estimates were calculated by computing average caustic position change for small changes in $s_{\rm b}$ and $s_{\rm p}$ and then converting these to $\Omega_{\rm E}$ estimates using \autoref{sdotratio}. The good agreement between numerical and analytical values is therefore only indicative of the validity of Equations~\ref{rabdots} and \ref{rabbdots} and not an endorsement of the validity of the analytic expression for $\Omega_{\rm E}$.

\subsection{General Orbits}

The effect of orbital motion on binary microlenses can be thought of as having rotational and separational components~\citep[e.g.,][]{Penny2011-om}. We have explored both of these components for circumbinary planets. In both cases the asymptotic behaviour of the orbital motion has the same scalings, and so it is reasonable to assume that these scalings apply in general. This also implies that despite the binary star causing some motion of the planetary caustics, except in certain cases, caustic motion will usually be dominated by the motion caused by the planet's orbital motion. It is worth noting however, that the orbital motion of the binary may instead be detectable as a so-called rapidly rotating lens~\citep{Penny2011-rrl,Nucita2014,Guo2015}.

\vspace{12pt}

\section{Will Circumbinary Systems Be Recognizable?}\label{fractionaldetectability}

In \autoref{caustics} we have discussed how certain features of circumbinary planet system caustics are unequivocally due to triple (or more) body lenses. However, it would be nice to be able to quantify \emph{how} detectable triple lens features will be in microlensing events, or at least how frequently easily recognizable triple-lens features might occur.

Properly assessing the detectability of triple-lens features would require simulating circumbinary lightcurves and fitting them with double-lens models to see if the triple-lens interpretation provides a significantly better fit. Similar efforts have been undertaken by \citet{Penny2011-om} and \citep{Zhu2014-fitting}, but it is extremely difficult to perform the fits on a large scale. Instead, we take  simpler approach as a first attempt at assessing whether circumbinary planets are recognizable as such. 

We restrict ourselves to mircolensing events with resonant caustic crossings, because these caustics are large and caustics crossings are less likely to be smeared together by finite source effects. We make the assumption that the lens would be recognizable as a triple lens if the lightcurve of the ABb lens has more caustic crossings than it would if the lightcurve were generated by the same trajectory through \emph{either} the AB or Ab lenses. We then assume that detailed modelling of the event would be able to distinguish between the various triple lens scenarios (e.g., circumbinary or circumprimary) as has been possible so far in caustic-crossing triple lens events~\citep{Gould2014}. This criterion has an implicit assumption: while our criterion only checks ABb caustics against AB and Ab caustics in the same configuration, it is possible that an AB or Ab system in a \emph{different} configuration could produce the same number of caustic crossings seen in the ABb system, however it is unlikely that it could produce the full underlying lightcurve.

\begin{figure}
\includegraphics[width=\columnwidth]{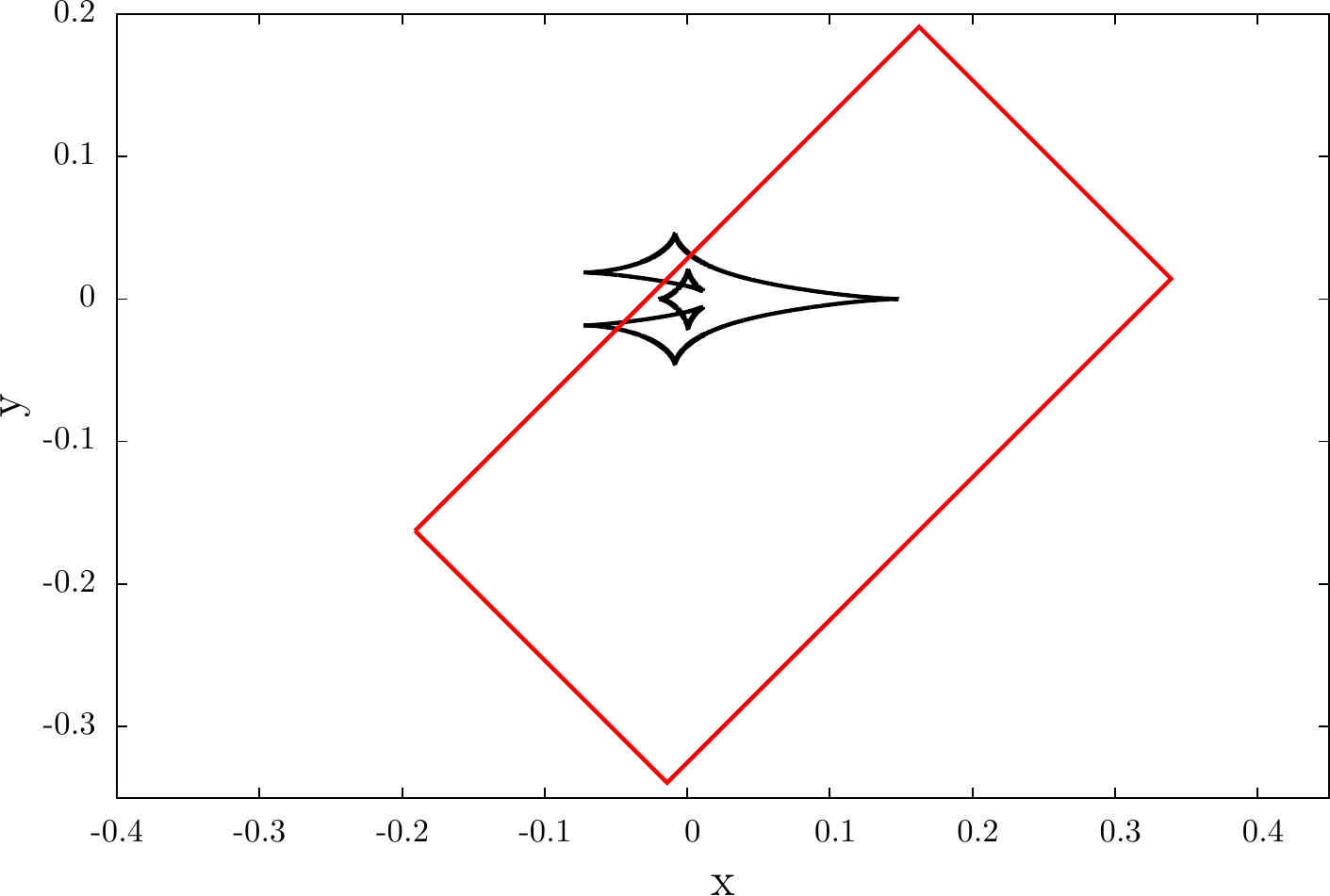}
\includegraphics[width=\columnwidth]{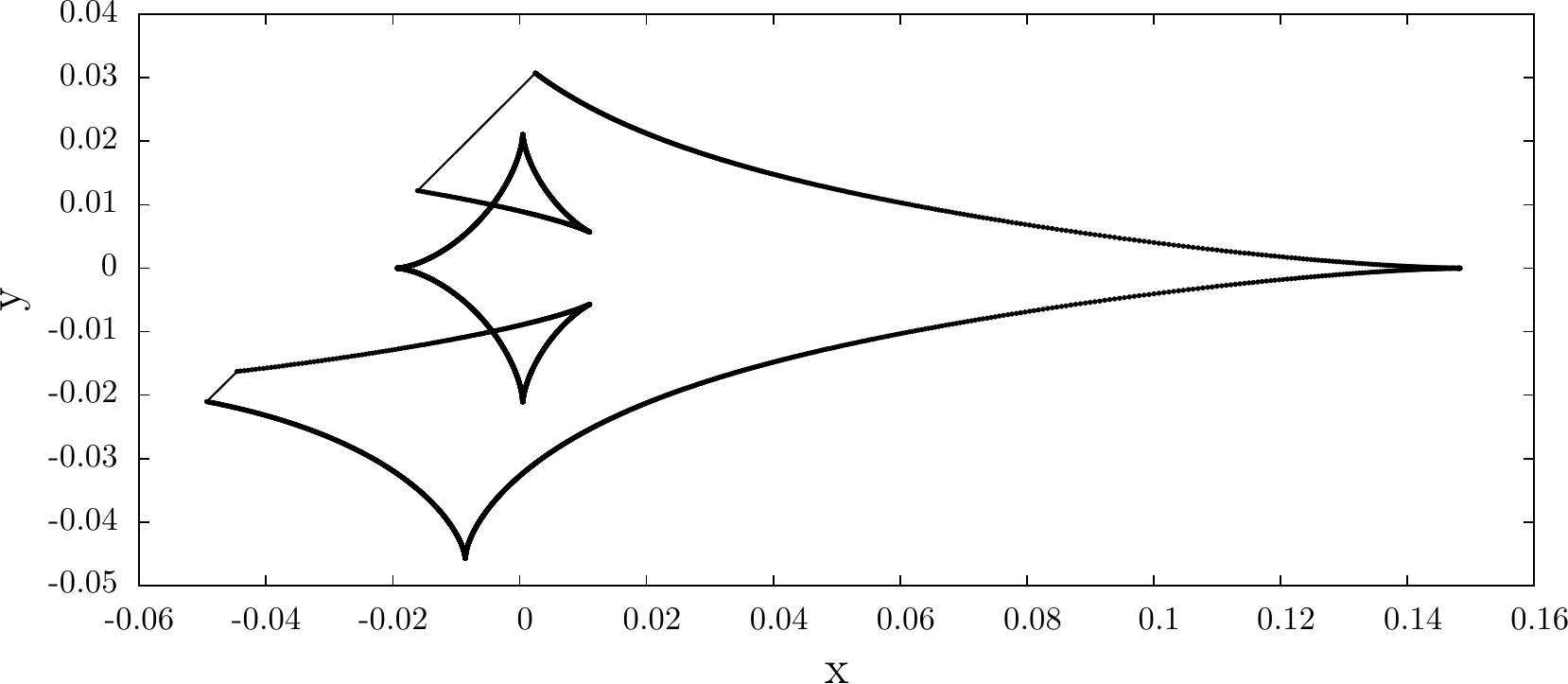}
\caption{\emph{Top panel}: an example caustic polygon and source trajectory polygon used for clipping. \emph{Bottom panel}: the resulting polygon after clipping.}
\label{polygon_clipping}
\end{figure}

To compute the number of caustic crossings for a given source trajectory and set of caustics, we first compute the caustics with a uniform sampling in the parameter $\psi$ (see \autoref{criticalcurvesoln}, the uniform sampling ensures that the cusps of the caustic are well sampled) and join together the $2N$ branches into closed polygons. For each source trajectory considered we then construct another polygon, with two vertices at opposite ends of a source trajectory that is guaranteed to start and end beyond all possible caustics and two more vertices again outside of all the caustics but slightly displaced from the two on the source trajectory such that the new polygon is a parallelogram or rectangle. We then \emph{clip} each caustic polygon with the source trajectory polygon, and search through the vertices of the resulting clipped polygon for any that lie on the source trajectory. These vertices are the points at which the source trajectory crosses the caustic. \autoref{polygon_clipping} shows an example caustic polygon, source trajectory polygon and the resulting clipped polygon. To perform the clipping, we used the intersection method of {\sc GPC} \citep[the General Polygon Clipping library][]{gpc}.\footnote{\url{http://www.cs.man.ac.uk/~toby/alan/software/}} This method was significantly faster than solving the lens equation along the source trajectory and searching for changes in the number of images.

We define the fractional circumbinary detectability as the fraction of caustic crossing source trajectories that produce a larger number of caustic crossings for the ABb caustic than both the AB and Ab caustics. For this to work it is important that the definition of the Einstein radius and origin be chosen so that the ABb caustics correspond exactly to the AB and Ab caustics when $s_{\rm p}$ or $q_{\rm b}$ and $s_{\rm p}$ or $q_{\rm p}$ are set to zero, respectively, as explained in \autoref{parameterspace}. To compute the fraction of trajectories with more caustic crossings we set up a grid of impact parameters $u_{\mathrm{0}}$ and trajectory angles $\beta$. To set the grid spacing, for one caustic configuration we kept doubling the number of samples in each dimension until the fractional detectability stabilized. This resulted in a grid with 80 $\beta$ samples and 320 $u_{\mathrm{0}}$ samples, for a total of 25,600 trajectories on our grid. 

\begin{figure*}
\includegraphics[width=\textwidth]{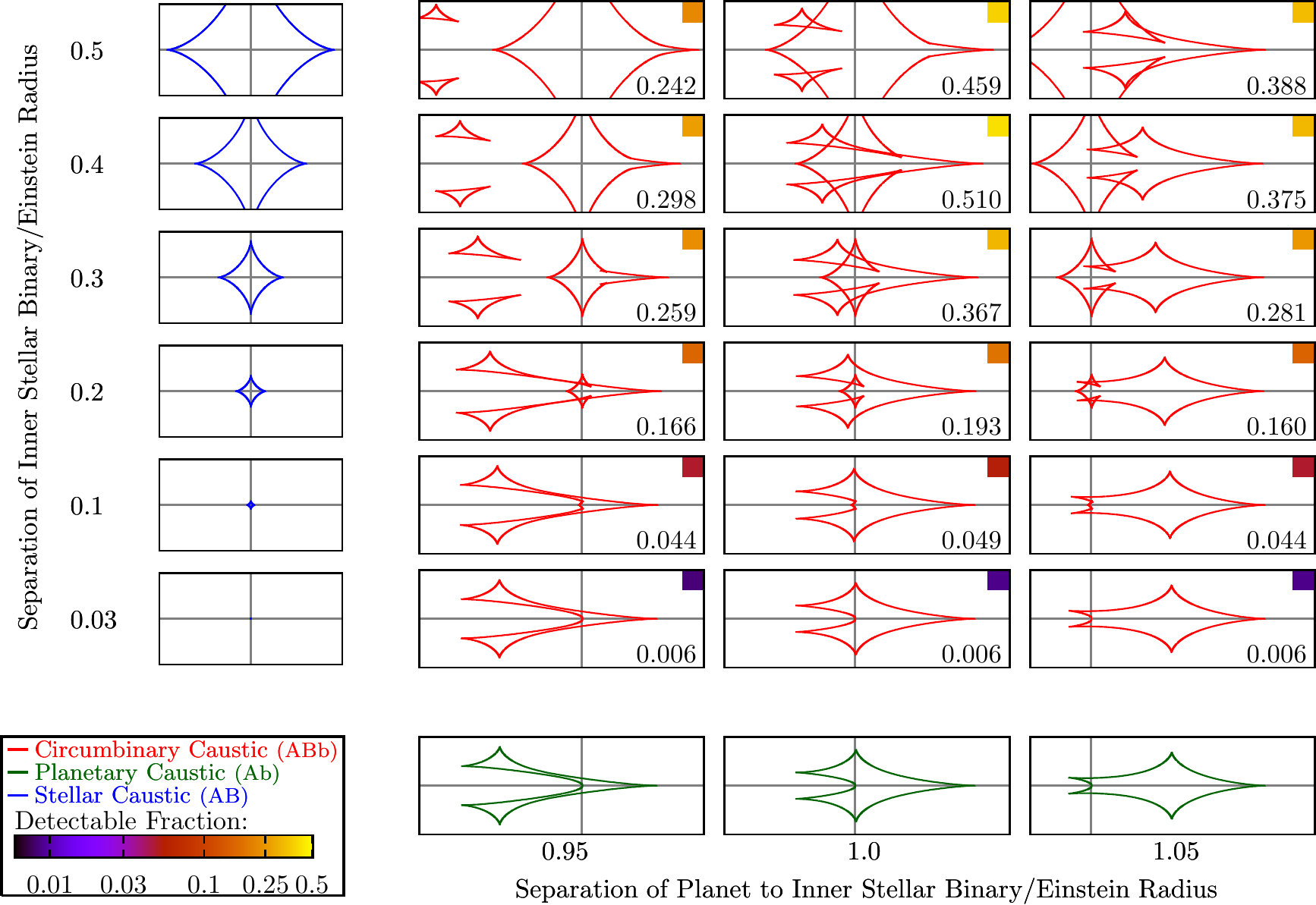}
\caption{Caustic structures of circumbinary systems and the ``detectable fraction'' as a function of $s_{\rm p}$ and $s_{\rm b}$ for $q_{\rm b} = 1$ and $q_{\rm p}=10^{-3}$. Detectable fraction is the fraction of caustic crossing trajectories where the number of caustic crossings is larger for the ABb caustic than either the AB and Ab caustics. At each point on the main $3\times6$ grid we plot the circumbinary ABb caustics for the $s_{\rm b}$ and $s_{\rm p}$ grid values; $s_{\rm b}$ increases vertically and $s_{\rm p}$ increases from left to right. We also plot the corresponding AB and Ab caustics in miniplots along the axes of the grid. In the top and bottom right corner of each mini-plot on the main grid we also show the detectable fraction as a number and a color-coded square.}
\label{fractional_detectability_figure}
\end{figure*}

For a planetary mass ratio $q_{\rm p}=10^{-3}$ and stellar binary mass ratio $q_{\rm b}=1$ we computed the circumbinary fractional detectability over a grid of 6 values of $s_{\mathrm{b}}$ between $0.03$ and $0.5$ \citep[note that for face-on orbits the systems with largest $s_{\rm b}$ relative to $s_{\rm p}\approx 1$ will be unstable, based on the stability criterion of][but that projection in non-face on systems can cause such large ratios of $s_{\rm b}$ to $s_{\rm p}$]{Holman1999}, 3 values of $s_{\mathrm{p}}$, and 18 values of $\phi_{\mathrm{b}}$. We noticed that the angle of the stellar binary, $\phi_{\mathrm{b}}$, did not affect the fraction significantly, so we averaged the fraction detected over the 18 angles. This left us with 18 circumbinary systems (6 $s_{\mathrm{b}}$, 3 $s_{\mathrm{p}}$) for which we obtained a fraction of trajectories that would be detected. \autoref{fractional_detectability_figure} shows the caustic structures of these 18 circumbinary systems, the caustics of the corresponding double-lens approximations, and the fraction of trajectories that would lead to a detected circumbinary system. 

\begin{figure*}
\includegraphics[width=0.495\textwidth]{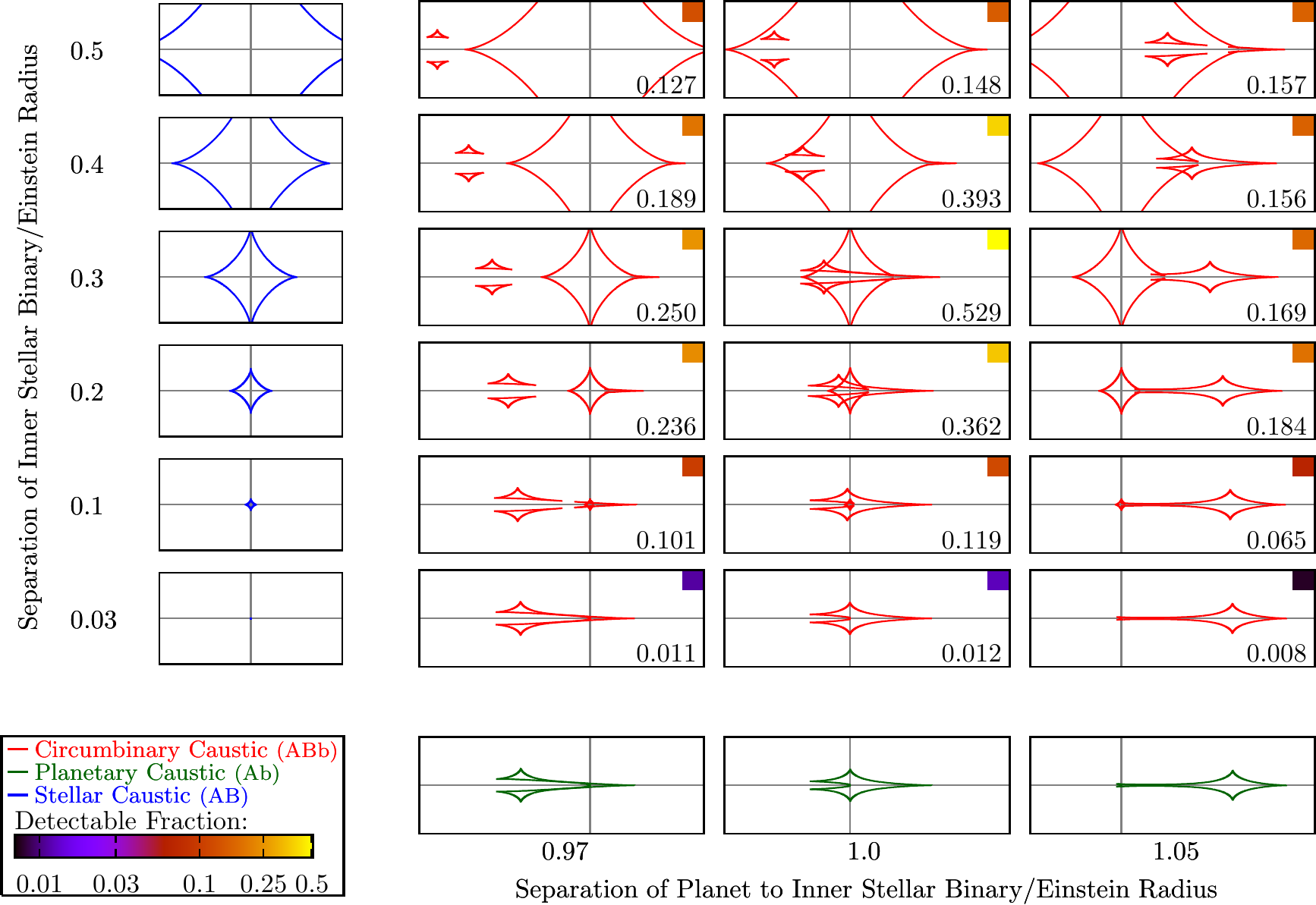}
\hspace{2pt}
\includegraphics[width=0.495\textwidth]{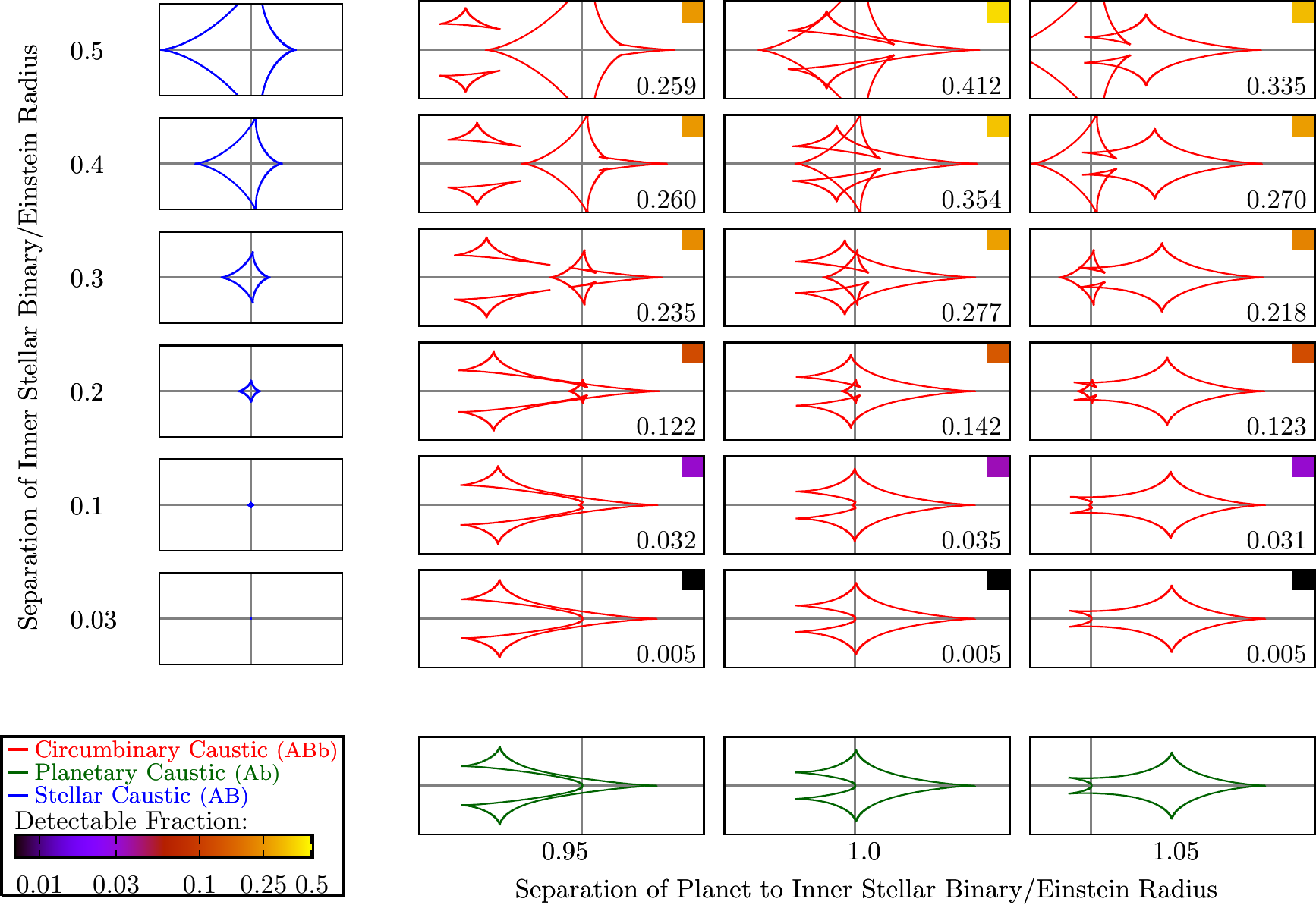}
\caption{As \autoref{fractional_detectability_figure} but the grid on the left has $q_{\rm p}=10^{-4}$ and the grid on the right has $q_{\rm b}=0.3$.}
\label{fractional_detectability_figure2}
\end{figure*}

As can be seen from the plot, the fractional detectability is largest when the stellar caustic and the planetary caustic are of similar size (i.e., when $s_{\rm b}\approx 0.4$ for this planetary mass ratio). In the regime where one caustic is larger than the other, the larger caustic dominates and there are fewer trajectories that would be detected as obviously circumbinary. For $s_{\rm b}\ge 0.1$ the fraction of trajectories that led to recognizable circumbinary systems was between 0.05 and 0.5. We measure similar fractional detectabilities for two more grids shown in \autoref{fractional_detectability_figure2}: one changes $q_{\rm p}$ to $10^{-4}$ and the other changes $q_{\rm b}$ to 0.3, compared to the first grid. This suggests that circumbinary planets will be detectable in a significant fraction of planetary microlensing events involving a resonant caustic over a wide range of planetary and stellar binary mass ratios and a reasonable range of binary semimajor axes.

\section{Discussion}\label{discussion}

Our work has only begun to scratch the surface of possible theoretical investigations of circumbinary planets. We have shown that the caustics of circumbinaries can yield to analytic investigations, and provide a general, if slightly cumbersome expression for the position of the circumbinary caustics in \autoref{planetarycausticslocation}, as well as a simplified expression for the special case of an equal mass binary host. Such analytic expressions are more than mere intellectual curiosities. They can play an important role in the analysis of microlensing events, significantly constraining the parameter space of potential solutions that need to be explored~\citep[e.g.][]{Poleski2014}. Our analytic expression for the caustic positions will be useful for efficiently exploring the degeneracy between the binary angle and separations of the binary and planet that will result from a caustic trajectory that encounters the planetary caustics in a circumbinary lens. We are sure that there are more analytic results to be discovered for circumbinary lenses that can further aid the interpretation of microlensing events in the future.

There is also significant room to improve on our estimates of the detectability of circumbinary planets. We considered only resonant caustics, and used a proxy for detectability that does not completely describe all the ways in which circumbinary planet lenses may be recognized as triple lenses. Specifically, we have only counted caustic crossings, but there will be many features in the magnification patterns of circumbinary lenses that will also reveal them to be triple lenses, such as spurs of higher magnification near to cusps \emph{internal} to caustics. We therefore expect our estimates of detectability for a given lens to be on the low side. This expectation is reinforced by the findings of \citet{Gould2014} that in circum\emph{primary} triple lens planetary events, the planet's effect on the central caustic that was dominated by the wide binary star was measureable despite there being no obvious features of triple-ness. However, our detection criterion also has a drawback: it does not say anything about how unique the lightcurve features will be to circumbinary planetary lenses amongst the various possible families of multi-body lenses. We expect that only detailed modeling analyses of a large number of triple lens lightcurves, both observed and simulated will reveal the importance of confusion between different triple lens solutions.

The prospects for detecting circumbinary planets via microlensing appear to be good. Observationally, the recent discoveries of \citet{Poleski2014} and \citet{Gould2014} have shown that the binarity of host stars can be identified in planetary microlensing events. \kepler's circumbinaries appear to be on the edge of what is detectable by microlensing, and with just slightly wider orbits of both the binary and the planet, systems similar to those found using \kepler\ can be found with microlensing. Our investigation, together with that of \citet{Han2008-tatooine} has shown that circumbinary planets can have interesting caustics with obvious triple-lens features. 

Given our optimism, we can ask why have no circumbinaries been found by microlensing yet? The answer probably comes down to both technique and time. Until the advent of the latest generation of high-cadence microlensing surveys, microlensing planet detections relied on follow-up observations by networks of small telescopes. These networks only have the resources to monitor a small fraction of ongoing microlensing events, and as such their targets must be prioritized. This lead to a situation where as soon as a microlensing event showed signs of being a binary star, e.g., a strong, large central caustic crossing, the follow-up networks had an incentive to stop observing the event as it would be less likely to yield a planet detection. This strategy will have undoubtedly reduced the detectability of planets in binary systems. However, the high-cadence surveys have been operating for a while now, and have found two circumprimary events~\citep{Poleski2014, Gould2014}; should we have expected them to have found circumbinary events by now too? Maybe not; we have shown that although there is a range of circumbinary parameter space that is accessible to microlensing, it is potentially limited by stability and simultaneous detectability of both components of the binary. This can limit the parameter space for circumbinary planets more than for circumprimary planets in binary systems with wide orbits. This is because the circumbinary host can only be detected over maybe half a dex of separations, but wide binaries can be detected over all separations, provided the source trajectory passes by both components of the binary~\citep[e.g.,][]{Poleski2014}. However, current microlensing surveys including MOA-II, OGLE-IV and KMTNet~\citep{Udalski2015-ogleiv,Sako2007,Henderson2014-kmt}, and those planned for the future, especially {\it Euclid} and {\it WFIRST} \citep{Penny2013, Spergel2015}, will vastly increase the number of planets found by microlensing and overcome the smaller parameter space for circumbinary planets. It is also worth noting that there are several existing microlensing events with caustic crossings for which double-lens solutions have not been found.

\section{Conclusion}\label{conclusions}

In this work we have focused on the caustic structure of circumbinary systems. Understanding the caustics is crucial to interpreting the lightcurves of circumbinary microlensing events. Our animations allowed us to explore the phenomenology of circumbinary caustics over a wide parameter space. Circumbinary planetary caustics can contain a number of features not seen in single-planet single-star caustics, including swallowtails and butterflies. When the planetary caustics are large, these features can be large and comparable to the size of the caustic, but are smaller in the central caustics of close and wide planets, and do not appear in the planetary caustics of close and wide lenses. Instead, the planetary caustics resemble those of the double lens Ab system, but are displaced by some amount depending on the stellar binary's parameters. Insights gained from the animations allowed us to derive a semi-empirical expression to describe this displacement. 

Our animations also act as crude simulations of orbital motion. Using our new analytic expression for the position of the caustics, we investigated the relative importance of the planetary and binary orbital motion to the motion of the caustics. We found that in all but a few exceptional cases, motion of the caustics is dominated by the orbital motion of the planet compared to the binary star. 

Finally, we investigated how detectable triple-lens features will be in circumbinary microlensing events with resonant caustic crossings. We found that a significant fraction ($5$-$50$~percent) of such events with binary projected separations $s_{\rm b}=0.1$--$0.5$ show differing numbers of caustic crossings in the circumbinary lens than in a single star planet host lens, which we use as a proxy for circumbinary detectability.


\vspace{12pt}
We thank Calen B. Henderson for his suggestions. We also thank the anonymous referee whose suggestions greatly increased the clarity of the paper. We acknowledge the support of the Summer Undergraduate Research Program (SURP) at the Ohio State University and NSF CAREER Grant AST - 1056524. This work was performed in part under contract with the California Institute of Technology (Caltech)/Jet Propulsion Laboratory (JPL) funded by NASA through the Sagan Fellowship Program executed by the NASA Exoplanet Science Institute.

\bibliographystyle{mn2e}
\bibliography{libraryshort,apj-jour}

\clearpage

\end{document}